\documentclass[a4paper,11pt]{article}
\pdfoutput=1 

\usepackage{jcappub} 

\title{Real space lensing reconstruction using cosmic microwave background polarization}

\author[a, b,1]{Heather Prince,\note{Corresponding author.}}
\author[b, c]{Kavilan Moodley,}
\author[d,b]{Jethro Ridl}
\author[e,b]{and Martin Bucher}

\affiliation[a]{Department of Astrophysical Sciences, Princeton University,\\
Peyton Hall, 4 Ivy Lane, Princeton, NJ, USA 08544}  
\affiliation[b]{Astrophysics and Cosmology Research Unit, \\ 
University of KwaZulu-Natal, Durban, 4041, South Africa}
\affiliation[c]{National Institute for Theoretical Physics,\\
University of KwaZulu-Natal, Durban, 4041, South Africa}
\affiliation[d]{Max-Planck-Institut f{\"u}r Extraterrestrische Physik,\\
 Giessenbachstra\ss e, D-85748 Garching, Germany}
\affiliation[e]{APC, AstroParticule et Cosmologie,\\ 
Universit\'e Paris Diderot, CNRS/IN2P3, CEA/lrfu, Observatoire de Paris, Sorbonne Paris Cit\'e, 10, rue Alice Domon et L\'eonie Duquet, 75205 Paris Cedex 13, France}

\emailAdd{heatherp@princeton.edu}
\emailAdd{moodleyk41@ukzn.ac.za}
\emailAdd{jridl@mpe.mpg.de}
\emailAdd{bucher@apc.univ-paris7.fr}

\abstract{We develop a method of reconstructing the lensing field from lensed CMB temperature and polarization maps in real space as an alternative to the harmonic space estimators currently in use by extending an existing real space lensing estimator for temperature to polarization. Real space estimators have the advantage of being local in nature and they are thus equipped to deal with the nonuniform sky coverage, especially galactic cuts and point source excisions, found in experimental data. 
We characterize some of the properties and limitations of these estimators and test them on simulated maps with Planck, AdvACT and CMB-S4 noise. We show that the reconstructions for large-scale lensing fields are accurate, and that the polarization reconstructions improve on those from CMB temperature maps for future experiments as expected. High-fidelity lensing maps can be reconstructed with futuristic experiments like CMB-S4.}

\begin{document}
\maketitle
\flushbottom

\section{Introduction}
\label{sec:intro}
The cosmic microwave background (CMB) anisotropy pattern on the last scattering surface is distorted by the gravitational field of the inhomogeneous matter density between the last scattering surface (at $z\approx 1100$) and us ($z=0$).
Thus gravitational lensing encodes information about the matter fluctuations in the CMB maps observed by our telescopes.  Apart from the obvious applications to constraining the matter density and the amplitude of the matter power spectrum, lensing is useful for breaking degeneracies between parameters that affect the CMB power spectrum in the same way. For example, measuring the lensing potential allows us to differentiate between curvature and dark energy for non-flat models of the universe \citep{Stompor1999}. Lensing can also improve constraints on dark energy models and neutrino masses, improving on the constraints from the lensed CMB power spectrum \citep{dePutter2009, Benoit-Levy2012, Allison2015Neutrinos}. Lensing probes the distribution of matter and thus serves as a tool to study massive objects such as galaxy clusters  \citep{Seljak2000, Dodelson2004, Holder2004, Baxter2015}, and can be especially useful in constraining cluster masses for high redshift objects with no background galaxies, where galaxy lensing cannot be used \citep{Hu&DeDeo&Vale2007}.

 Additionally, cross correlations of CMB lensing reconstructions with other tracers of the dark matter distribution can be used to study the clustering properties of the tracers, such as their bias with respect to the underlying dark matter field \citep{Sherwin2012, Holder2013, Pearson2014, Kuntz2015, Bianchini2015, Allison2015}.  Cross-correlation of a foreground lens plane with lensing from two different background source planes (such as the CMB at $z \approx 1100$ and background sources for galaxy lensing at much lower redshifts) probes the geometrical distances to different redshifts, and can thus be used to constrain dark energy parameters \citep{Jain2003, Bernstein2004, Zhang2005, Das2009, Miyatake2017}. Combining results from CMB and galaxy lensing reconstructions \citep{Hand2015, Liu2015} also allows us to constrain the amplitude of the density fluctuations probed by both types of lensing.

The primordial CMB is very close to Gaussian \citep{Planck2015nonGaussianity}. It is therefore fully described by its angular power spectrum, and different harmonic modes are independent.  Gravitational lensing introduces non-Gaussianities into the lensed CMB \citep{Bernardeau1997, Zaldarriaga2000, Hu2001} and induces correlations between different harmonic modes. Because these off-diagonal correlations are proportional to the lensing potential, quadratic estimators can be applied to the CMB temperature and polarization data in harmonic space to reconstruct the lensing potential. This technique for reconstructing the lensing potential using harmonic space quadratic estimators was first proposed in ref. \citep{Zaldarriaga1999} and further developed in refs. \citep{Seljak1999, Hu&Okamoto2002, Okamoto&Hu2003, Cooray2003}. Alternative methods of reconstructing CMB lensing using likelihood-based methods have also been explored \citep{Hirata2003a, Hirata2003b}.

Initial lensing detections were achieved by cross-correlating with data from the Wilkinson Microwave Anisotropy Probe (WMAP) satellite \citep{Smith2007, Hirata2008}. The lensing power spectrum was then measured using CMB temperature data observed by the Atacama Cosmology Telescope (ACT) \citep{Das2011} and the South Pole Telescope (SPT) \citep{vanEngelen2012}. Polarization lensing was first detected in cross correlation by the South Pole Telescope Polarimeter (SPTpol) and \textit{Herschel} \citep{Hanson2013}. The lensing power spectrum from CMB polarization data was first measured by the POLARBEAR collaboration \citep{Ade2014}.  Gravitational lensing has now been detected in both temperature and polarization data by various CMB experiments such as the Atacama Cosmology Telescope Polarimeter (ACTPol) \citep{Sherwin2016}, SPTpol \citep{Story2015} and the BICEP2/Keck collaborations \citep{BICEP&Keck2016}. The most significant lensing detection to date is the Planck satellite's 40$\sigma$ measurement, achieved using a combination of temperature and polarization data \citep{Planck2015Lensing}. In practice, the most widely used estimator has been the quadratic estimator of ref. \citep{Hu&Okamoto2002}. 

In this paper, we derive real space estimators that reconstruct the lensing convergence and shear directly from CMB temperature and polarization maps, without the need to transform the data into harmonic space. The real space approach can be helpful when analyzing experimental data, as it makes use of local estimators which can easily cope with non uniform sky coverage and pixels that have been removed from CMB maps, whereas the ubiquitous harmonic space estimators \citep[e.g.][]{Hu&Okamoto2002} implicitly require uniform full sky coverage to work without adaptations. The estimators presented here take the form of the convolution of a CMB temperature or  polarization map with a real space kernel (having a limited spatial range) multiplied by another CMB map. This paper extends previous work \citep{Bucher2012}, which focused on temperature lensing estimators, to include polarization estimators.

 As polarization maps of the microwave sky improve, reconstructions of the CMB gravitational lensing potential will rely almost exclusively on polarization data, with temperature data contributing very little additional information \citep{Hu&Okamoto2002}. This is because for the unlensed polarization anisotropies the polarization B mode power in the CMB due to `scalar' primordial cosmological perturbations, which are the only ones we have observed so far \citep{Ade2015}, vanishes. Therefore, whatever B modes we may detect are entirely due to gravitational lensing (neglecting astrophysical foregrounds and primordial tensor perturbations for now). Because the B modes arising from the scalar mode vanish, the measured B mode from lensing is not polluted by `cosmic variance,' quite unlike the situation for the temperature anisotropies, where it is difficult to separate the lensed CMB signal from statistical fluctuations in the primordial 
signal due to cosmic variance. In principle, measuring the B mode polarization perfectly, with no detector noise or polarized B mode foreground contaminants, under the assumption that primordial tensor modes are absent, would allow us to reconstruct the gravitational lensing potential perfectly, with no error at all. This is to be contrasted with the lensing potential reconstruction using the temperature maps, where even perfect measurements would lead to error in the reconstructed lensing potential because of cosmic variance. When polarization measurements on small scales reach a quality such that the measurement error becomes comparable to the lensed B mode signal, the lensing reconstruction from polarization will surpass that available from exploiting the temperature maps.

The properties of the lensing estimators depend on the experimental noise. We compare results from a Planck-like experiment \citep{Planck2015Spectra} to current experiments with better resolution and noise properties, such as the Advanced Atacama Cosmology Telescope Polarimeter (AdvACT) \citep{Henderson2015}, and future planned `Stage 4' experiments such as CMB-S4 \citep{S4_2016}. 
	The polarization reconstruction from $EB$ is expected to have a lower reconstruction noise than the $TT$ estimator for experiments with noise below $4-5 \mu \text{K arcmin}$ \citep{Hu&Okamoto2002,Kesdenetal2003, S4_2016}. Consequently, the polarization estimators become very important for the CMB-S4 experiment.
	Ground-based experiments such as AdvACT and CMB-S4 do not observe the full sky and so a local treatment is useful, especially if the survey strategy includes deep observations in relatively small patches of the sky.

The paper is organized as follows. We derive temperature and polarization lensing estimators in section \ref{sec:setup} and explain their implementation in real space in section \ref{sec:RS_implementation}. In section \ref{sec:RS_from_HS} we show that the real space estimators can be derived from the standard harmonic space estimator. We present an alternative derivation based on lensed correlation functions, resulting in expressions for estimating the lensing fields from lensed $T$, $Q$, and $U$ maps in section \ref{sec:corr_fn}. We then describe the multiplicative bias that affects our reconstructions and how to correct for it in section \ref{sec:formfactor}. In section  \ref{sec:reconstructions} we apply the real space estimators to simulated maps to produce reconstructed maps of the lensing fields. We conclude with a discussion of our results and an outlook on future work in section \ref{sec:discussion}.

\section{Constructing temperature and polarization estimators}
\label{sec:setup}

Gravitational lensing deflects CMB photons, remapping them on the sky by the deflection angle $\boldsymbol{\alpha}$ which can be expressed as the gradient of the lensing potential $\psi$.
The lensed temperature and Stokes linear polarization parameters (denoted by a tilde) can be related to the unlensed temperature or polarization by \citep{Blanchard1987, Bernardeau1997, Zaldarriaga&Seljak1998}
\begin{equation}
\begin{split}
\tilde{T}(\boldsymbol{x}')=T(\boldsymbol{x}) \,,
\qquad
\tilde{Q}(\boldsymbol{x}')=Q(\boldsymbol{x}) \,,	
\qquad
\tilde{U}(\boldsymbol{x}')=U(\boldsymbol{x}) \,,
\end{split}
\end{equation}
 where the positions in the unlensed and lensed sky are related, for small deflections, by the linear transformation $\boldsymbol{x}=\boldsymbol{S}\boldsymbol{x}'$ with $\boldsymbol{S}=e^{\boldsymbol{\kappa}}$. 
The deformation tensor $\boldsymbol{\kappa}$ describes the distortion of the primordial CMB anisotropies due to gravitational lensing, and is given by the derivative of the deflection angle $\boldsymbol{\alpha}$
 	\begin{equation}
 	{\kappa}_{ij}=\nabla_i\alpha_j=\nabla_i\nabla_j\psi=
	 \begin{pmatrix}
	  \kappa_0+\gamma_+ & \gamma_\times \\
	  \gamma_\times & \kappa_0-\gamma_+
 	 \end{pmatrix}.
	 \label{eq:deformation_tensor}
 	\end{equation}

The convergence $\kappa_0=\frac{1}{2} \boldsymbol{\nabla}\cdot \boldsymbol{\alpha}= \frac{1}{2} \nabla^2 \psi$ determines the shape-preserving expansion or contraction of a source due to lensing. The components of the shear, given by $\gamma_+= \frac{1}{2} (\frac{\partial^2 \psi}{\partial x^2} - \frac{\partial^2 \psi}{\partial y^2}  )$ and $\gamma_\times = \frac{\partial^2 \psi}{\partial x \partial y }$, determine the distortion of the shape of the source along different axes due to the tidal gravitational field. The lensing convergence and the two components of the shear field are locally well defined, while the lensing potential $\psi$ and the deflection angle $\boldsymbol{\alpha}=\boldsymbol{\nabla} \psi$ are both ambiguous: $\psi$ is indistinguishable from $\psi+ \text{(constant)}$, and the vector field $\boldsymbol{\alpha}$ is indistinguishable from a translation (in the flat sky approximation) or a rotation (when we take sky curvature into account) of itself, since we do not have access to a map of the unlensed CMB. We will thus focus in this paper on developing local estimators for the lensing convergence and shear fields in map space.
	
\begin{figure}[h]
 \centering
\includegraphics[width=0.5\textwidth]{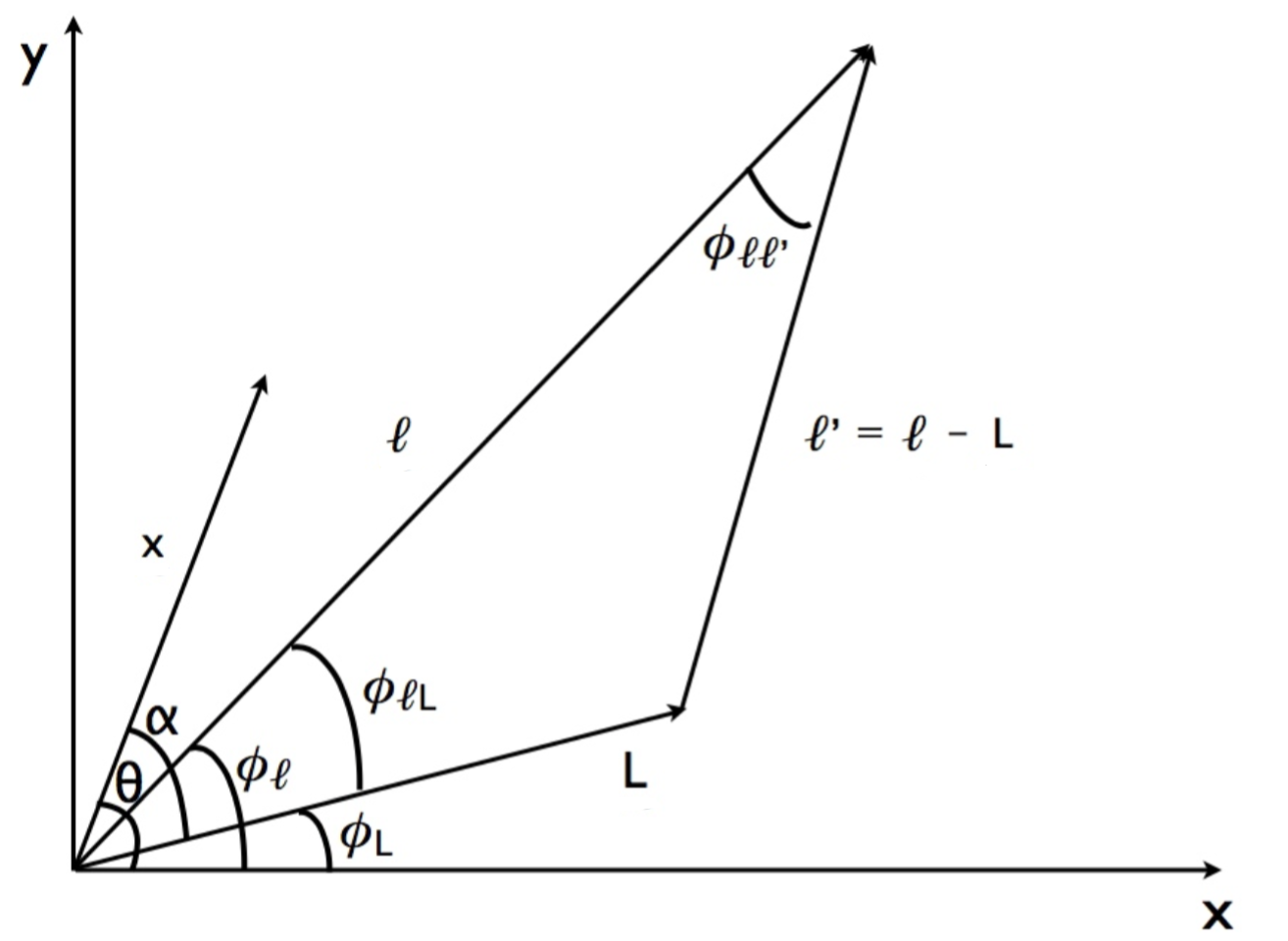}
\caption{Coordinate system in map and Fourier space. The length of $\boldsymbol{L}$ has been exaggerated to make the diagram more readable.  }
\label{fig:coords}
\end{figure}

Gravitational lensing introduces statistical anisotropies to the CMB, resulting in nonzero off-diagonal elements in the lensed CMB covariance matrix \citep{Hu2000, Hanson2010}. Thus the lensing field couples initially uncorrelated harmonic CMB modes. The approximation used below in deriving the real space lensing estimators focuses on large-scale lensing modes ($\kappa_0$, $\gamma_+$ and $\gamma_\times$), which act on small-scale CMB anisotropies. This corresponds in Fourier space to CMB modes with large angular wavevectors $\boldsymbol{\ell}$ and $\boldsymbol{\ell}'$, which are coupled by a much smaller lensing wavevector $\boldsymbol{L}=\boldsymbol{\ell}-\boldsymbol{\ell}'$, resulting in a `squeezed triangle' as seen in figure \ref{fig:coords}. This is a reasonable approximation because small scale anisotropies contribute most of the statistical information about lensing \citep{Seljak1996, Metcalf&Silk1997}, and the lensing potential peaks at fairly low $L$.  
However, the squeezed triangle approximation will result in a biased reconstruction of small-scale lensing modes for the estimators derived here. This effect is discussed in more detail in section \ref{sec:formfactor}.

We consider only weak lensing of the CMB, in which case the deflection angle is small. The lensing effect is thus perturbatively small, and we neglect all terms of second order or higher in the lensing fields. The polarization basis propagated along the perturbed photon path (in direction $\boldsymbol{x}'$ on the sky) is rotated slightly with respect to the basis propagated in the direction $\boldsymbol{x}$ on the sky. This rotation angle is less than $\sim$ 1 arcminute provided the two bases are related by parallel transport along a spherical geodesic  \citep{Lewis&Challinor2006}. Thus the effect of this rotation can be neglected.


\subsection{Lensed angular power spectra}

	  The Fourier transforms of the unlensed and lensed temperature maps in the flat sky approximation are found by integrating over areas $\mathcal{A}$ and $\mathcal{A}'$ respectively to obtain 
	  \begin{align}
	  T(\boldsymbol{\ell})&=\frac{1}{\sqrt{\mathcal{A}}} \int d^2\boldsymbol{x}\, T(\boldsymbol{x}) e^{-i\boldsymbol{\ell}\cdot\boldsymbol{x}} \nonumber \\ 
	  \tilde{T}(\boldsymbol{\ell})&=\frac{1}{\sqrt{\mathcal{A}'}} \int d^2\boldsymbol{x}\,' \,\tilde{T}(\boldsymbol{x}\,') e^{-i\boldsymbol{\ell}\cdot\boldsymbol{x}'}  =\text{det}^{-\frac{1}{2}}(\boldsymbol{S}) T(\boldsymbol{S}^{-1}\boldsymbol{\ell}),
	  \label{eq:temp}
 	  \end{align}
 	  where the primed and unprimed coordinates are related by the Jacobian  $J$ of the coordinate transformation $\boldsymbol{S}^{-1}$ and $\det{J}={1}/{\det{\boldsymbol{S}}}$.
 
 	 In Fourier space linear polarization can be separated into  $E$ modes, which are polarized parallel or perpendicular to the angular wavevector, and $B$ modes, which are polarized at a $45^\circ$ angle to the $E$ modes \citep{Hu&White1997}. Thus the polarization $E$ and $B$ modes in Fourier space are related to the Stokes $Q$ and $U$ parameters in real space by
 	 \begin{align}
  	 E(\boldsymbol{\ell})\pm 
   	 iB(\boldsymbol{\ell})&=\frac{1}{\sqrt{\mathcal{A}}} \int d^2\boldsymbol{x} \left(Q(\boldsymbol{x})+iU(\boldsymbol{x}) \right) e^{-i\boldsymbol{\ell}\cdot \boldsymbol{x}} e^{\mp 2 i \phi_{\ell}} \nonumber \\  
   	\tilde{E}(\boldsymbol{\ell})\pm i\tilde{B}(\boldsymbol{\ell}) &= \frac{1}{\sqrt{\mathcal{A}'}} \int d^2\boldsymbol{x}\,'  \left(\tilde{Q}(\boldsymbol{x}\,')+i\tilde{U}(\boldsymbol{x}\,') \right)  e^{-i\boldsymbol{\ell}\cdot \boldsymbol{x}'}   e^{\mp 2 i \phi_{\ell}}   \nonumber \\  
  	 & =\text{det}^{-\frac{1}{2}}(\boldsymbol{S}) \left(E(\boldsymbol{\ell}') \pm iB(\boldsymbol{\ell}')\right) e^{\mp 2 i (\phi_{\ell}-\phi_{\ell'})}, \label{eq:X}
  	 \end{align}
   	 where the lensed wavevector is related to the unlensed wavevector by $\boldsymbol{\ell}'=\boldsymbol{S}^{-1}\boldsymbol{\ell}$ and $\phi_{\ell'}$ is the polar angle of $\boldsymbol{\ell}'$.  
 Rewriting eq. (\ref{eq:X}) by expressing the angular factor as
        	 \begin{equation}
     e^{\mp 2 i (\phi_{\ell}-\phi_{\ell'})} = 1 \mp 2i[\gamma_\times \cos(2\phi_{\ell})-\gamma_+ \sin(2\phi_{\ell})],
     \label{eq:e}
     \end{equation}
      which is an approximation valid to first order in the lensing fields, we obtain
     \begin{equation}
     \tilde{E}(\boldsymbol{\ell})=\text{det}^{-\frac{1}{2}}(\boldsymbol{S})[E(\boldsymbol{\ell}')+ 
     2(\gamma_\times \cos(2\phi_{\ell})-\gamma_+  \sin(2\phi_{\ell}))B(\boldsymbol{\ell}')] \nonumber
     \end{equation}
     \begin{equation}
     \tilde{B}(\boldsymbol{\ell})=\text{det}^{-\frac{1}{2}}(\boldsymbol{S})[B(\boldsymbol{\ell}') -
     2(\gamma_\times \cos(2\phi_{\ell})-\gamma_+ \sin(2\phi_{\ell}))E(\boldsymbol{\ell}')],
     \end{equation}
     where $\boldsymbol{\ell}'=\boldsymbol{S}^{-1}\boldsymbol{\ell}$ as before. 	
A calculation of the determinant keeping only first-order terms yields det$^{-1}(\boldsymbol{S})=1-2\kappa_0$.

 	It is now straightforward to relate the lensed angular power spectra and cross spectra of the CMB temperature and $E$ and $B$ mode polarization to the primordial spectra $C_\ell^{XY}$ defined by
\begin{equation}
\langle X^*(\vec{\ell})Y(\vec{\ell'}) \rangle=(2\pi)^2\delta_D^2(\vec{\ell}-\vec{\ell'})C^{XY}_\ell,
\end{equation}
where $XY$ varies over any of the relevant map combinations, namely $TT$, $EE$, $BB$, $EB$, $TE$ or $TB$.\footnote{In the expectation value above, we average over realizations of the unlensed CMB temperature and polarization maps, keeping the lensing potential fixed, which is the convention used hereafter.} The primordial $B$ mode power spectrum $C^{BB}_{\ell}$ is expected to be a very small signal from gravitational waves, which can be neglected for our purposes, and the unlensed cross spectra $C^{EB}_{\ell}$ and $C^{TB}_{\ell}$ are zero by parity invariance. 
	For $XY=TT$, $EE$ and $TE$, we find the lensed angular power to be
 	\begin{align}
  	\tilde{C}^{XY}_{\ell}\,&=\,\text{det}^{-1}(\boldsymbol{S}) C^{XY}_{|\boldsymbol{S}^{-1}\boldsymbol{\ell}|} =\,(1-2\kappa_0)\left(C^{XY}_{\ell}+\mathcal{K} l\frac{dC^{XY}_{\ell}}{d\ell}\right) ,
	\end{align}
	where we have made use of the relation $\ell'\equiv|\boldsymbol{S}^{-1}\boldsymbol{\ell}|=\ell-\mathcal{K}\ell$ with
		\begin{equation}
 	\mathcal{K} \equiv \frac{\boldsymbol{\ell}\cdot \boldsymbol{\kappa}\cdot \boldsymbol{\ell}}{l^2}=\kappa_0+\gamma_+ \cos(2\phi_{\ell})+\gamma_\times \sin(2\phi_{\ell}).
 	\label{eq:defK}
 	\end{equation}
This gives the anisotropic lensed spectrum
	\begin{align}
 	\tilde{C}^{XY}_{\ell}\,&=\,C^{XY}_{\ell} -\kappa_0\left(\frac{dC^{XY}_{\ell}}{d\ln\ell}+2C^{XY}_{\ell}\right)-(\gamma_+ \cos(2\phi_{\ell})+\gamma_\times \sin(2\phi_{\ell}))\frac{dC^{XY}_{\ell}}{d\ln\ell} ,
	\label{eq:spec1}
	\end{align}
from which it can be seen that if the local spectral index $\beta\equiv \frac{d\ln C^{XY}_{\ell}}{d\ln{\ell}}$ has the value $\beta=-2,$ corresponding to a scale-invariant spectrum, the power spectrum $C^{XY}_{\ell}$ is invariant under pure dilatations, as expected. If $\beta=0$, corresponding to a white noise spectrum, the power spectrum is invariant under pure shear transformations  \citep{Bucher2012}.

The remaining lensed power spectra are given by:
 	\begin{equation}
 	\tilde{C}^{YB}_{\ell}=\langle \tilde{Y}^*(\boldsymbol{\ell})\tilde{B}(\boldsymbol{\ell}) \rangle=- 2 C^{YE}_{\ell}\left[\gamma_\times \cos(2\phi_{\ell})-\gamma_+ \sin(2\phi_{\ell})\right],
 	\end{equation}
where $Y$ is either $T$ or $E.$ In our approximation, the convergence has no effect on the lensed $TB$ and $EB$ power spectra, and thus we cannot use these spectra to reconstruct $\kappa_0$. This is because dilating $E$ mode polarization with a large-scale convergence field that is approximately constant over the sky does not convert any of the $E$ mode polarization pattern into $B$ modes, resulting in the lensed $TB$ and $EB$ spectra being zero. Shearing effects, however, mix polarization $E$ and $B$ modes, and we can thus use the lensed $TB$ and $EB$ combinations to reconstruct the shear. The lensed $TB$ and $EB$ spectra are only sensitive to derivatives of the unlensed spectra at higher order, unlike the lensed $TT, TE$ and $EE$ spectra.

 	 In the squeezed triangle approximation, when we neglect primordial $B$ modes, we find that the lensed $B$ mode spectrum vanishes. In reality $\tilde{C}^{BB}_{\ell}$ will be nonzero even if the unlensed spectrum is zero because $E$ mode polarization is converted to $B$ mode polarization under lensing \citep{Zaldarriaga&Seljak1998}. However, this effect is second order in $\gamma_+$ and $\gamma_\times$, and thus negligible in the linear and squeezed triangle approximations used here. We therefore do not consider estimators based on $\tilde{C}^{BB}_{\ell}$.

 	\subsection{Estimators}
\label{sec:estimators}	
		The squeezed triangle approximation, under which we assume a small lensing wavenumber $L$, corresponds to $\kappa_0$, $\gamma_+$ and $\gamma_\times$ varying slowly across the sky. We use our expressions for the lensed power spectra to find estimators for $\kappa_0$, $\gamma_+$ and $\gamma_\times$ in a region of area $\mathcal{A}$ over which these quantities are constant. We can think of this region as being a pixel in the map. The estimator can then be translated to all pixels in the map.
 
	 An ansatz for the $\kappa_0$ estimator, based on eq. (\ref{eq:spec1}), involves the weighted average of products of lensed maps:
	 \begin{equation}
	 \hat{\kappa}_0^{XY}=\frac{1}{N_{\hat{\kappa}_0}^{XY}} 
	 \int \frac{d^2\boldsymbol{\ell}}{(2 \pi)^2} \left(\tilde{X}^*(\boldsymbol{\ell})\tilde{Y}(\boldsymbol{\ell})-C_{\ell}^{XY}\right)g_{\hat{\kappa}_0}^{XY}(\ell),
 	\label{eq:kappaXY}
 	\end{equation} 
 	where the unlensed power spectrum $C^{XY}_{\ell}$ is subtracted from the observed one to isolate $\kappa_0,$ and  $XY=TT$, $EE$, or $TE$.
We choose the normalization constant $N_{\hat{\kappa}_0}^{XY}$ to make the estimator unbiased, i.e., $\langle \hat{\kappa}_0^{XY}\rangle=\kappa_0$. The weight function $g_{\hat{\kappa}_0}^{XY}$ is chosen to minimize the variance of the estimator.
	 The variance of the convergence estimator receives contributions from cosmic variance and CMB detector noise $n^{XY}(\ell),$ which depends on the resolution and sensitivity of the experiment, as discussed further in section \ref{sec:noise}.

\begin{table}[!tbp]
\footnotesize
\centering
{\def\arraystretch{2}\tabcolsep=10pt
\begin{tabular}{|c| c| c|}
\hline
Estimator & Weight Function $g(\ell)$ & Normalization $N$  \\ 
\hline

 $\hat{\kappa}_0^{TT}$ & $\left(\frac{d\ln C^{TT}_{\ell}}{d\ln\ell}+2\right) \frac{C^{TT}_{\ell}}{(C^{TT}_{\ell}+n^{TT}(\ell))^2}$ & $\int \frac{d^2\boldsymbol{\ell}}{(2 \pi)^2} g_{\hat{\kappa}_0}^{TT}(\ell)  \left(\frac{dC^{TT}_{\ell}}{d\ln\ell}+2C^{TT}_{\ell}\right)$ \\ 
 
 $\hat{\kappa}_0^{EE}$ & $\left(\frac{d\ln C^{EE}_{\ell}}{d\ln\ell}+2\right) \frac{C^{EE}_{\ell}}{(C^{EE}_{\ell}+n^{EE}(\ell))^2}$ & $\int \frac{d^2\boldsymbol{\ell}}{(2 \pi)^2} g_{\hat{\kappa}_0}^{EE}(\ell)  \left(\frac{dC^{EE}_{\ell}}{d\ln\ell}+2C^{EE}_{\ell}\right)$ \\ 

  $\hat{\kappa}_0^{TE}$ & $\left(\frac{dC^{TE}_{\ell}}{d\ln\ell}+2C^{TE}_{\ell}\right) \frac{1}{\left(\tilde{C}^{TE}_{\ell}\right)^2+\left(C^{TT}_{\ell}+n^{TT}(\ell)\right)\left(C^{EE}_{\ell}+n^{EE}(\ell)\right)}  $ & $ \int \frac{d^2\boldsymbol{\ell}}{(2 \pi)^2} g_{\hat{\kappa}_0}^{TE}(\ell) \left(\frac{dC^{TE}_{\ell}}{d\ln\ell}+2C^{TE}_{\ell}\right)$ \\ 

  \hline

   $\hat{\gamma}_{+,\times}^{TT}$ & $ \left(\frac{d \ln C^{TT}_{\ell}}{d\ln\ell}\right) \frac{C^{TT}_{\ell}}{(C^{TT}_{\ell}+n^{TT}_{\ell})^2}$ & $\frac{1}{2} \int \frac{d^2\boldsymbol{\ell}}{(2 \pi)^2}g_{\hat{\gamma}_+, \hat{\gamma}_\times}^{TT}(\ell)  \frac{dC^{TT}_{\ell}}{d\ln\ell}  $ \\ 
 
  $\hat{\gamma}_{+,\times}^{EE}$ & $  \left(\frac{d \ln C^{EE}_{\ell}}{d\ln\ell}\right) \frac{C^{EE}_{\ell}}{(C^{EE}_{\ell}+n^{EE}_{\ell})^2} $ & $\frac{1}{2} \int \frac{d^2\boldsymbol{\ell}}{(2 \pi)^2}g_{\hat{\gamma}_+, \hat{\gamma}_\times}^{EE}(\ell)  \frac{dC^{EE}_{\ell}}{d\ln\ell}  $ \\ 
 
 $\hat{\gamma}_{+,\times}^{TE}$ & $ \left(\frac{dC^{TE}_{\ell}}{d\ln\ell}\right) \frac{1}{(\tilde{C}^{TE}_{\ell})^2+(C^{TT}_{\ell}+n^{TT}(\ell))(C^{EE}_{\ell}+n^{EE}(\ell))}$ & $\frac{1}{2}\int \frac{d^2\boldsymbol{\ell}}{(2 \pi)^2} g_{\hat{\gamma}_+, \hat{\gamma}_\times}^{TE}(\ell)\frac{dC^{TE}_{\ell}}{d\ln\ell}$ \\ 

  $\hat{\gamma}_{+,\times}^{EB}$ & $\frac{C^{EE}_{\ell}}{(C^{EE}_{\ell}+n^{EE}(\ell))(n^{BB}(\ell))} $ & $\int \frac{d^2\boldsymbol{\ell}}{(2 \pi)^2} g_{\hat{\gamma}_+, \hat{\gamma}_\times}^{EB}(\ell) C^{EE}_{\ell} $ \\ 

   $\hat{\gamma}_{+,\times}^{TB}$ & $\frac{C^{TE}_{\ell}}{(C^{TT}_{\ell}+n^{TT}(\ell))(n^{BB}(\ell))} $ & $\int \frac{d^2\boldsymbol{\ell}}{(2 \pi)^2} g_{\hat{\gamma}_+, \hat{\gamma}_\times}^{TB}(\ell)  C^{TE}_{\ell} $ \\

\hline
\end{tabular}
}
\caption{The normalization and weight functions for the convergence and shear estimators given in eqs. (\ref{eq:kappaXY}) and (\ref{eq:gammaXY}).}
\label{table:estimators}
\end{table}

Estimators for $\gamma_+$ and $\gamma_\times$ can be found by multiplying the lensed power spectra by $\cos(2\phi_{\ell})$ and $\sin(2\phi_{\ell})$ before averaging, to isolate the shear components.
We obtain
	\begin{equation}
	\begin{Bmatrix}{\hat{\gamma}_+^{XY}}\\{\hat{\gamma}_\times^{XY}}\end{Bmatrix}=\frac{1}{N_{\hat{\gamma}_+, \hat{\gamma}_\times}^{XY}} \int \frac{d^2\boldsymbol{\ell}}{(2 \pi)^2} g_{\hat{\gamma}_+, \hat{\gamma}_\times}^{XY}(\ell)
	 \begin{Bmatrix}{\cos(2\phi_{\ell})}\\{\sin(2\phi_{\ell})} \end{Bmatrix}\tilde{X}^*(\boldsymbol{\ell})\tilde{Y}(\boldsymbol{\ell})
	 \label{eq:gammaXY}
	\end{equation}
for all map combinations $XY.$ The weight functions and normalization constants for the different estimators are shown in table \ref{table:estimators}. There is no $EB$ or $TB$ convergence estimator because these combinations are not affected by the large-scale lensing convergence, as discussed above.

These quadratic estimators for the lensing convergence and shear are unbiased and of minimum variance in the squeezed triangle approximation. The estimators in this section demonstrate how we can reconstruct the lensing convergence and shear in one pixel using temperature and polarization data in harmonic space. In the following section we extend these to estimators that act directly in map space.

\section{Implementation in real space}
\label{sec:RS_implementation}
The estimators for the lensing convergence and shear formulated in the previous section reconstruct the lensing fields from harmonic space CMB temperature and polarization fields. Using harmonic space quantities derived from experimental data can be challenging because of sky cuts, point source excisions, and nonuniform sky coverage \citep{Hanson2009}, 
so it is helpful to formulate lensing estimators that act on (real space) CMB maps. Since the estimators are formulated in terms of a product in harmonic space, the estimators defined in the map space  correspond to a convolution of CMB maps. 

For the convergence estimator we have 
	\begin{align}
	\hat{\kappa}_0^{XY}(\boldsymbol{x}_0)&= \int d ^2\boldsymbol{x} \tilde{X}(\boldsymbol{x}_0-\boldsymbol{x})  \int d^2 \boldsymbol{x}\,^\prime \tilde{Y}(\boldsymbol{x}\,^\prime) K_{\hat{\kappa}_0}^{XY}  (|\boldsymbol{x}_0-\boldsymbol{x}-\boldsymbol{x}\,^\prime|)	\nonumber \\
	&=\int d^2 \boldsymbol{x}  \tilde{X}(\boldsymbol{x}_0-\boldsymbol{x})(K_{\hat{\kappa}_0}^{XY} \circ \tilde{Y})(\boldsymbol{x}_0-\boldsymbol{x}) \, , 
	\label{eq:RSimplementation}
	\end{align}
	where $XY=TT$, $EE$ or $TE,$ and $\circ$ denotes convolution. We have neglected the contribution from the unlensed component for now but include it in the final expression below. The lensing kernel in real space $K_{\hat{\kappa}_0}^{XY}( \boldsymbol{x})$ is  the inverse Fourier transform of the weight function $g_{\hat{\kappa}_0}^{XY}(\ell)$ normalized by $N_{\hat{\kappa}_0}^{XY}$
		\begin{equation}
	K_{\hat{\kappa}_0}^{XY}(x) =\frac{1}{ N_{\hat{\kappa}_0}^{XY}}  \int \frac{d^2\ell}{(2\pi)^2} e^{i\boldsymbol{\ell}\cdot\boldsymbol{x}}g_{\hat{\kappa}_0}^{XY}(\ell)
	=\frac{1}{ N_{\hat{\kappa}_0}^{XY}}  \int \frac{d\ell}{2\pi} \ell J_0(\ell x) g_{\hat{\kappa}_0}^{XY}(\ell),
	\label{eq:KapKernel}
	\end{equation}
where $J_0$ is a zeroth order Bessel function of the first kind.

The kernel $K_{\hat{\kappa}_0}^{XY}(x)$ peaks at small angular scales, as discussed further in section \ref{sec:kernels}, which means that the central pixel $\boldsymbol{x}_0$ contains the greatest signal and integrating over other pixels makes the reconstruction noisier (also see \citep{Zhu2015}). This allows us to approximate the convergence at this point in the map as $\hat{\kappa}_0^{XY}(\boldsymbol{x}_0) \approx \tilde{X}(\boldsymbol{x}_0)(K_{\hat{\kappa}_0}^{XY} \circ \tilde{Y})(\boldsymbol{x}_0)$. Translating this to different points in the map yields the real space estimator
	\begin{equation}
	\hat{\kappa}_0^{XY}(\boldsymbol{x})=\tilde{X}(\boldsymbol{x})(K_{\hat{\kappa}_0}^{XY} \circ \tilde{Y})(\boldsymbol{x})-\langle X(\boldsymbol{x})(K_{\hat{\kappa}_0}^{YY} \circ Y)(\boldsymbol{x}) \rangle_{\text{unlensed}}\,, 
	\label{eq:KappaEst}
	\end{equation}
	where we have now removed the contribution to the convergence from the unlensed component.
	
Unlike the convergence kernel, the kernel for the shear estimator is anisotropic, reflecting the anisotropic effect of the shear field. We can rewrite the shear estimators in map space as
	 	\begin{equation}
 	\begin{Bmatrix}{\hat{\gamma}_+^{XY}(\boldsymbol{x})}\\{\hat{\gamma}_\times^{XY}(\boldsymbol{x})}\end{Bmatrix}=\tilde{X}(\boldsymbol{x})(K_{\tiny{\begin{Bmatrix}{\hat{\gamma}_+}\\{\hat{\gamma}_\times}\end{Bmatrix}}}^{XY} \circ \tilde{Y})(\boldsymbol{x}),
	\label{eq:GammaEst}
 	\end{equation}
	where the kernels are given by
	 	\begin{equation}
 	K_{\tiny{\begin{Bmatrix}{\hat{\gamma}_+}\\{\hat{\gamma}_\times}\end{Bmatrix}}}^{XY}(\boldsymbol{x}) = 
    {K_{\hat{\gamma}_+, \hat{\gamma}_\times}^{XY}(x) }
    \begin{Bmatrix}{\cos{2\theta(\boldsymbol{x})}}\\{\sin{2\theta(\boldsymbol{x})}}\end{Bmatrix},
	\label{eq:ShearKernel2}
 	\end{equation}
	for all combinations $XY.$ Here $\theta(\boldsymbol{x})$ is a map of the polar angle of $\boldsymbol{x}$ and 
 	\begin{equation}
 	 	K_{\hat{\gamma}_+, \hat{\gamma}_\times}^{XY}(x) =\frac{1}{N_{\hat{\gamma}_+, \hat{\gamma}_\times}^{XY} } \int_0^\infty \frac{ d\ell}{2\pi} \ell \,J_2(\ell x) g_{\hat{\gamma}_+, \hat{\gamma}_\times}^{XY}(\ell),
	\label{eq:ShearKernel}
 	\end{equation}
 	where $J_2$ is the second order Bessel function of the first kind.
     
By Fourier transforming the above expressions,
we can generalize the harmonic space expressions for the real-space convergence and shear estimators presented in the previous section to nonzero wavelengths $L>0,$ giving
\begin{equation}
	\hat{\kappa}_0^{XY}(\boldsymbol{L})=\tilde{X}(\boldsymbol{L})\circ (K_{\hat{\kappa}_0}^{XY}  \tilde{Y})(\boldsymbol{L})-\langle X(\boldsymbol{L})\circ(K_{\hat{\kappa}_0}^{YY}  Y)(\boldsymbol{L}) \rangle_{\text{unlensed}}\, 
	\label{eq:KappaEst}
	\end{equation}
and
\begin{equation}
 	\begin{Bmatrix}{\hat{\gamma}_+^{XY}(\boldsymbol{L})}\\{\hat{\gamma}_\times^{XY}(\boldsymbol{L})}\end{Bmatrix}=\tilde{X}(\boldsymbol{L})\circ (K_{\tiny{\begin{Bmatrix}{\hat{\gamma}_+}\\{\hat{\gamma}_\times}\end{Bmatrix}}}^{XY}  \tilde{Y})(\boldsymbol{L}).
	\label{eq:GammaEst}
 	\end{equation}
The corresponding harmonic space kernels can be obtained by Fourier transforming the real space kernels and using the fact that $e^{-i{\bf L}\cdot {\bf x}} = J_0(Lx) + 2 \sum_{n=1}^{\infty} (-i)^n J_n(Lx) \cos(n\alpha),$ where $\cos(n\alpha) = \cos(n\phi_L) \cos(n\theta) +  \sin(n\phi_L) \sin(n\theta).$ We find that the convergence harmonic space kernel is 
\begin{equation}
	K_{\hat{\kappa}_0}^{XY}(\boldsymbol{L}) =\frac{g_{\hat{\kappa}_0}^{XY}(L)}{N_{\hat{\kappa}_0}^{XY}} ,
	\label{eq:KapKernel2}
	\end{equation}
    for $XY$=$TT$, $TE$, and $EE,$ and the
shear harmonic space kernels are
\begin{equation}
 	K_{\tiny{\begin{Bmatrix}{\hat{\gamma}_+}\\{\hat{\gamma}_\times}\end{Bmatrix}}}^{XY}(\boldsymbol{L}) = \frac{g_{\hat{\gamma}_+, \hat{\gamma}_\times}^{XY}(L)}{N_{\hat{\gamma}_+, \hat{\gamma}_\times}^{XY}} \begin{Bmatrix}{\cos{2\phi_L}}\\{\sin{2\phi_L}}\end{Bmatrix}
	\label{eq:ShearKernel2}
 	\end{equation}
for all map combinations $XY,$ where $g_{\hat{\kappa}_0}^{XY}$ and $g_{\hat{\gamma}_+, \hat{\gamma}_\times}^{XY}$ are given in table \ref{table:estimators}. 
 
Lensing reconstruction using the convergence and shear estimators presented above is efficient if the reconstruction kernels are limited in spatial extent, so that most of the lensing information comes from small scales. We devote the rest of this section to examining which CMB scales contain the most information about the lensing signal as well as exploring the shape and spatial extent of the lensing reconstruction kernels.

\subsection{Noise and foregrounds}
\label{sec:noise}
The quality of the lensing reconstruction is limited by the experimental noise in the maps. The optimal real-space lensing reconstruction kernel depends on the noise spectrum, downweighting noisy scales in the reconstruction. For a Gaussian beam, the noise spectrum is given in terms of the pixel noise and the beam window function as
\begin{equation}
n(\ell)=\sigma_p^2 e^{\ell^2\theta_{beam}^2}
\end{equation}
where $\theta_{beam}$ is the beam width and $\sigma_p$ is the detector noise in a pixel of side $\theta_{beam}=\theta_{FWHM}/\sqrt{8 \ln 2}$.
 We will study real-space lensing reconstructions for the Planck 143 GHz channel \citep{Planck2015HFI} in comparison to Advanced ACTPol's 150 GHz channel \citep{Henderson2015} and a future CMB-S4 mission \citep{S4_2016}. 
 The specifications for these experiments are shown in table \ref{table:CMBspecs}.
 
 \begin{table}[tbp]
\centering
\begin{tabular}{|c| c c c c|}
\hline
Experiment & Channel & Beam Size $ \theta_{FWHM}$ & Temperature Sensitivity  &  Polarization Sensitivity \\ 
\hline
Planck & 143 GHz &7.3 arcmin & 33  $\mu$K-arcmin & 70 $\mu$K-arcmin  \\ 
AdvACT & 150 GHz &1.4 arcmin & 7 $\mu$K-arcmin & 10 $\mu$K-arcmin  \\ 
CMB-S4 & 150 GHz &1 arcmin & 1 $\mu$K-arcmin & 1.4 $\mu$K-arcmin  \\ 
\hline
\end{tabular}
\caption{Planck, Advanced ACTPol and CMB-S4 experimental specifications}
\label{table:CMBspecs}
\end{table}

In order to apply these estimators to experimental data, foregrounds must be taken into account. 
The foregrounds begin to dominate the CMB temperature signal at $\ell \sim 3000$. Thus for higher resolution experiments such as AdvACT, much of the apparent small scale signal will actually be from foregrounds, which will act as an additional source of noise in our reconstruction. To mitigate this, we add the predicted foreground signal to our experimental noise term when we calculate the lensing reconstruction kernels, thus cutting off the kernels on scales that are dominated by foregrounds. We include the thermal and kinetic Sunyaev-Zeldovich (tSZ and kSZ) effects, the Poisson and clustered cosmic infrared background (CIB) components, radio galaxies, galactic dust emission and the cross correlation between tSZ and the CIB in our foreground template, according to the prescription in ref. \citep{Dunkley2013}. We only make use of the angular power spectrum of the foregrounds to downweight modes by how contaminated they are, and do not take into account the non-Gaussianity of these foregrounds, which biases the lensing reconstruction \citep{vanEngelen2014}. For polarized foregrounds, we consider only polarized point sources \citep{Naess2014ACTPol}, assuming that the power spectrum has the same flat shape as the Poisson sources in temperature but a lower amplitude, as point sources are only partially polarized. We assume a characteristic fractional polarization of 10\%, which is probably an overestimate \citep{Tucci2012}.  A more complete foreground treatment should take into account the non-Gaussianity of the foregrounds, as well as the effect of foregrounds not included here such as the polarized emission from galactic dust. However, as the main focus of this paper is to develop and test real space estimators, the simplified treatment described above is sufficient for our purposes.


\subsection{Reconstruction kernels}
\label{sec:kernels}

\begin{figure}[!htb]
\centering
\includegraphics[width=\textwidth]{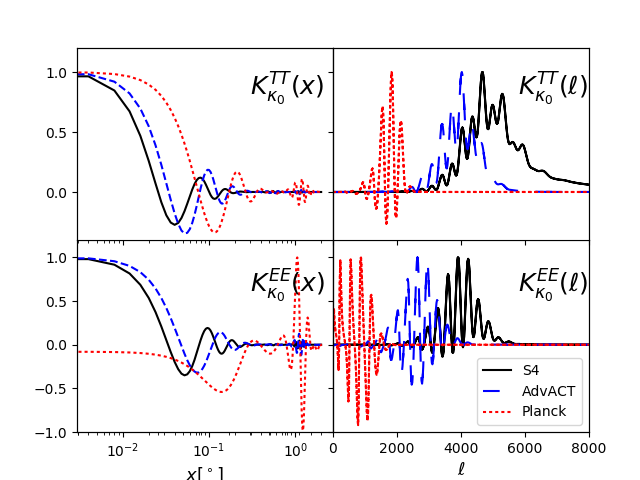}
\caption{TT (top row) and EE (bottom row) convergence lensing kernels in real space (left column) and in Fourier space (right column)}
\label{fig:kernelC}
\end{figure}

\begin{figure}[!htb]
\centering
\includegraphics[width=\textwidth]{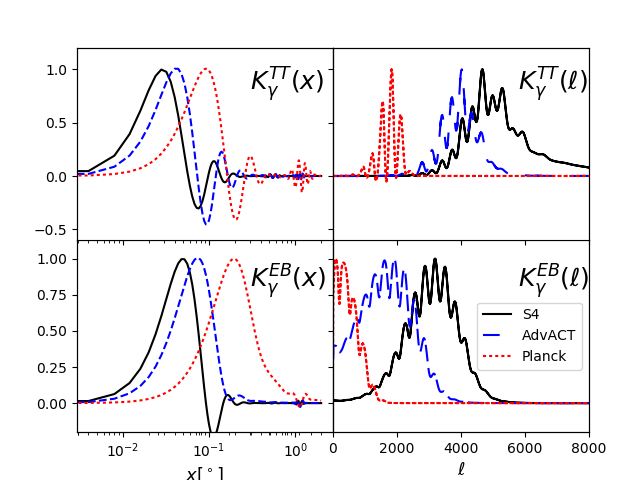}
\caption{TT (top row) and EB (bottom row) shear lensing kernels in real space (left column) and in Fourier space (right column) }
\label{fig:kernelS}
\end{figure}

\begin{figure}[!htb]
\centering
\includegraphics[width=\textwidth]{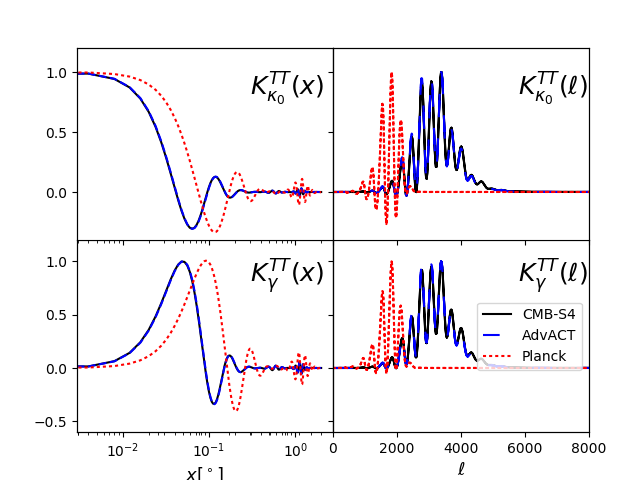}
\caption{TT convergence (top row) and shear (bottom row)  lensing kernels in real space (left column) and in Fourier space (right column) with foregrounds included in the noise term. The inclusion of foregrounds cuts off the kernels at a lower angular wavenumber. }
\label{fig:kernelCfgnd}
\end{figure}

We now study the shape of the lensing reconstruction kernels. As discussed earlier, the real-space filters need to be limited in spatial extent if we are to perform localized reconstructions on a small patch of the sky. The shapes of the lensing kernels in real space and harmonic space are shown in figure \ref{fig:kernelC} for the convergence estimators, and in figure \ref{fig:kernelS} for the shear estimators. We only show the $TT$ and $EE$ kernels for the convergence estimator, as the $TE$ kernel shape is qualitatively similar to $EE,$ and the $TT$ and $EB$ kernels for the shear estimators, as the $TE$ and $EE$ shear kernels are qualitatively similar to the $TT$ kernel and the $TB$ kernel similar to the $EB$ kernel.
The kernels for the temperature estimator have been presented previously \citep{Bucher2012} but the polarization kernels are shown here for the first time. The kernels in figures \ref{fig:kernelC} and  \ref{fig:kernelS} include detector noise but no foregrounds. The temperature kernels with foregrounds included in the noise term are shown in figure \ref{fig:kernelCfgnd}. The shape of the polarization kernels is not affected by including foregrounds because we have only taken into account Poisson sources in the polarized foregrounds, which means that the additional noise term is flat in harmonic space. The plots are normalized to peak at unity for ease of comparison between the experimental configurations, which have different absolute normalizations.

As claimed earlier, the estimator kernels are compact and all vanish beyond an angular scale of $2^\circ$. However, given the varying beam size and detector noise levels of CMB experiments considered here, there are expected differences between their respective kernels. The kernels for AdvACT peak at higher $\ell$ than those for Planck, which means that the lensing reconstruction from maps with AdvACT noise and resolution receives much of its signal-to-noise from small-scale CMB modes that are lensed by large-scale lensing fields. This is even more the case for CMB-S4 in the absence of foregrounds, whereas when foregrounds are included, the kernels for CMB-S4 are pushed to lower $\ell,$ closer to those for AdvACT but still more compact than the Planck kernels, which hardly change when foregrounds are included. 

The acoustic scale of $\delta \ell \approx 200$  imprinted on $K(\ell)$ as a harmonic modulation, manifests itself in the real space kernel at degree scales. This effect corresponds to lensing of the CMB acoustic feature in real space, a feature which is seen in CMB maps stacked on hot spots or cold spots \citep{Komatsu2011}, and indicates that lensing of CMB hot spots or cold spots is correlated with lensing of CMB photons approximately 1 degree away. This suggests that real-space lensing estimators need to operate on patches at least 1 degree wide to capture this signal, a requirement that is somewhat independent of the beam size. Even for very high resolution and low noise CMB experiments, the lensed acoustic feature in $K(x)$ is present at some level and contributes to lensing reconstruction. For Planck since the beam smears out the lensing information on very small scales the lensed acoustic signature at the $1^\circ$ scale is relatively more prominent than for AdvACT and CMB-S4, particularly for the EE spectrum.

The convergence real space kernel peaks at the origin with a  width inversely proportional to the width $\Delta \ell$ of the broadband envelope of the harmonic space kernel, which is set by a combination of the beam and detector noise relative to the CMB signal. The smaller beam and detector noise of AdvACT and CMB-S4 (with $\Delta \ell \approx 4000$) gives more weight to the kernel at smaller angular scales (below 3') compared to Planck.
The shear kernels share the same features as the convergence kernels, with the only major difference being that the real-space shear kernels peak slightly away from the origin. This is because the shear kernels result from an integration over $J_2(\ell x)$ (due to the $\cos 2 \theta$ angular dependence), which has a zero at the origin, in contrast to the convergence kernel, which is isotropic and picks out $J_0(\ell x)$, which does not vanish at the origin. Physically this makes sense because the shear is reconstructed from a distortion of the polar tangent vector, which has a length that goes to zero at the origin, while the convergence relies on a distortion of the radial tangent vector, with length independent of $x$.

Most of the weight in the real space lensing kernels shown above is contained within $2^\circ$, with the kernels for AdvACT peaking at smaller angular scales than those for Planck. As discussed in ref. \citep{Bucher2012}, we can choose to restrict the real space filter to angular scales smaller than $x_{max}$ and then compute the optimal filter for that truncation scale from the full kernel. The limited angular extent of the filters means that we can perform the convolutions for the real space estimator over small areas of the sky, allowing us to perform localized reconstructions of the lensing convergence and shear.

In this section we have considered the shape of the lensing kernels, with and without foregrounds, to highlight how foregrounds limit the extent to which lensing information can be extracted on small scales. In reality foregrounds must be accounted for in the reconstruction kernel.

 \subsection{Cumulative information}
 \label{sec:cum_info}

The lensing estimators developed above are unbiased and of minimum variance in the limit where the lensing fields are constant over the patch being reconstructed. 
To evaluate how well the estimators reconstruct lensing fields that vary over the patch, it is helpful to investigate which CMB scales contribute the most statistical weight to the lensing estimators. We study this by considering the cumulative information or signal-to-noise, which also indicates how well the different estimators perform relative to each other. 

The normalization factors $N_{\kappa_0}$ and $N_{\gamma}$ given in table \ref{table:estimators} measure the inverse variance of the convergence and shear estimators \citep{Hu&Okamoto2002, Lewis&Challinor2006}, or equivalently the signal-to-noise squared per unit solid angle per unit distortion of $\kappa_0$, $\gamma_+$ or $\gamma_\times$.  The CMB wavenumbers $\ell$ that dominate the integral dominate the lensing reconstruction. Moreover, the lensing reconstruction is accurate for lensing wavenumbers much smaller than these CMB wavenumbers, i.e., in the squeezed triangle approximation described in section \ref{sec:setup}.

\begin{figure}[!bht]
\begin{minipage}{0.49\linewidth}
  \centering
  \includegraphics[width=\textwidth]{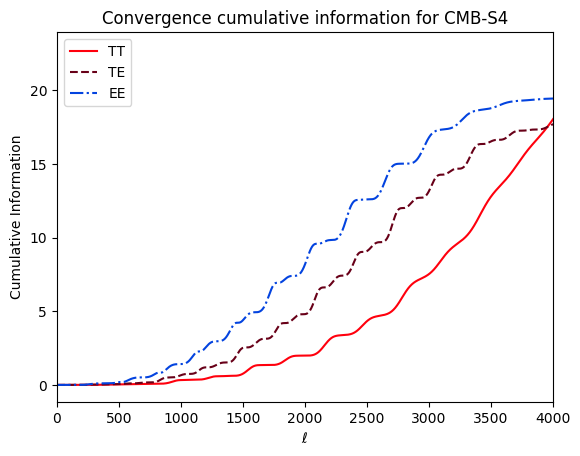}
  \end{minipage}
  \begin{minipage}{0.49\linewidth}
  \centering
  \includegraphics[width=\textwidth]{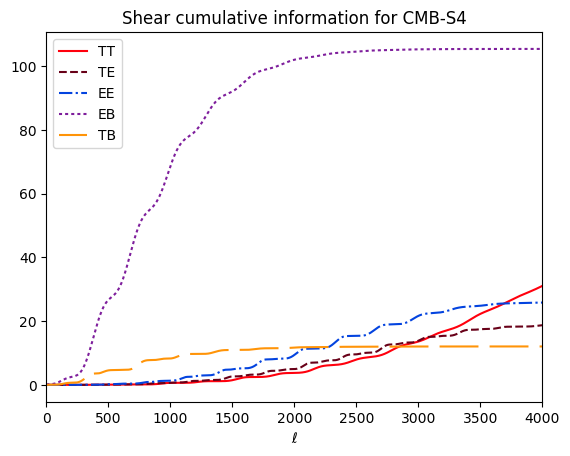}
  \end{minipage}

\begin{minipage}{0.49\linewidth}
  \centering
  \includegraphics[width=\textwidth]{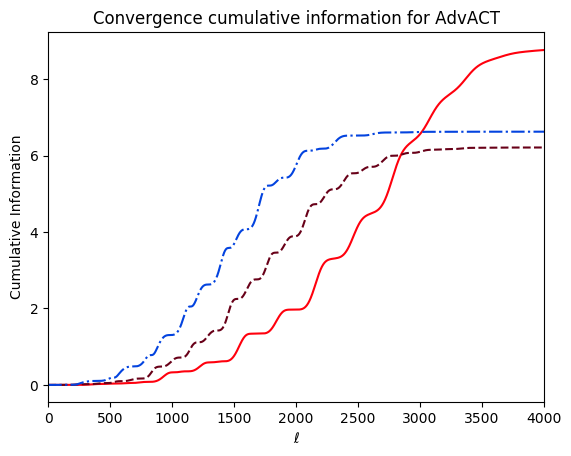}
  \end{minipage}
  \begin{minipage}{0.49\linewidth}
  \centering
  \includegraphics[width=\textwidth]{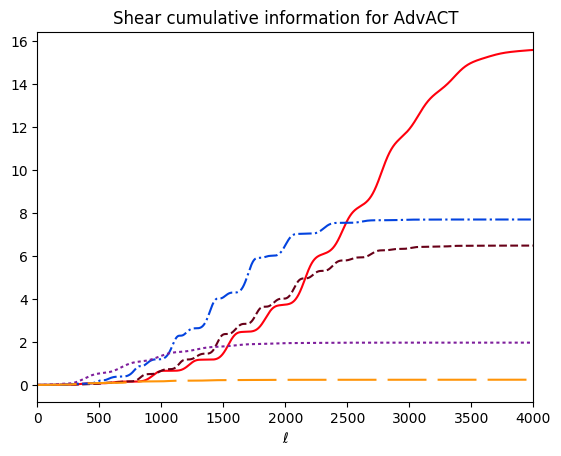}
  \end{minipage}
  
  \begin{minipage}{0.49\linewidth}
  \centering
  \includegraphics[width=\textwidth]{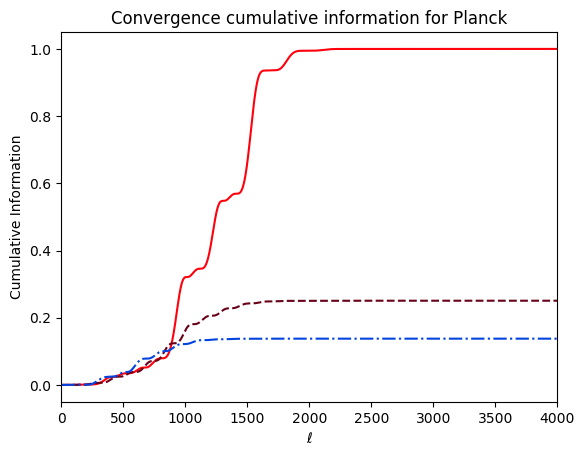}
  \end{minipage}
  \begin{minipage}{0.49\linewidth}
  \centering
  \includegraphics[width=\textwidth]{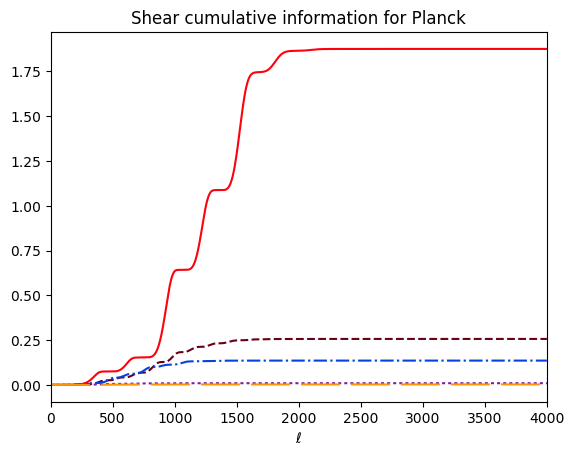}
  \end{minipage}
  \caption{Cumulative information for different temperature and polarization estimators, calculated using noise properties of  CMB-S4 (top) AdvACT (center) and Planck (bottom). Only the experimental noise is included in the noise term in the kernel: foregrounds are not taken into account.}
 \label{fig:cumulative_no_FG}
\end{figure}

\begin{figure}[!bht]
\begin{minipage}{0.49\linewidth}
  \centering
  \includegraphics[width=\textwidth]{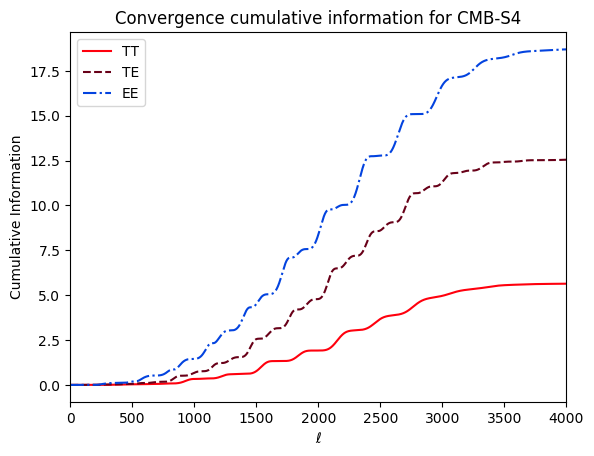}
  \end{minipage}
  \begin{minipage}{0.49\linewidth}
  \centering
  \includegraphics[width=\textwidth]{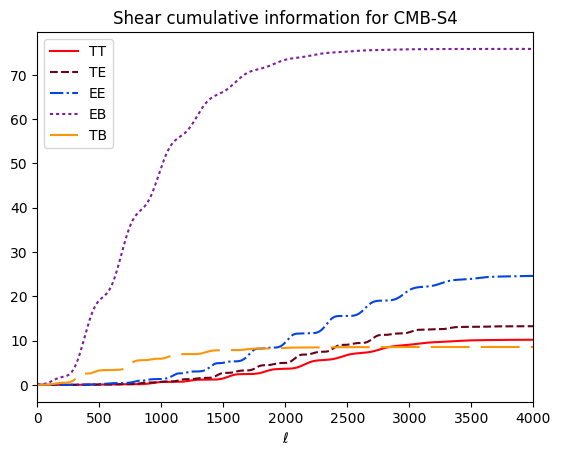}
  \end{minipage}

\begin{minipage}{0.49\linewidth}
  \centering
  \includegraphics[width=\textwidth]{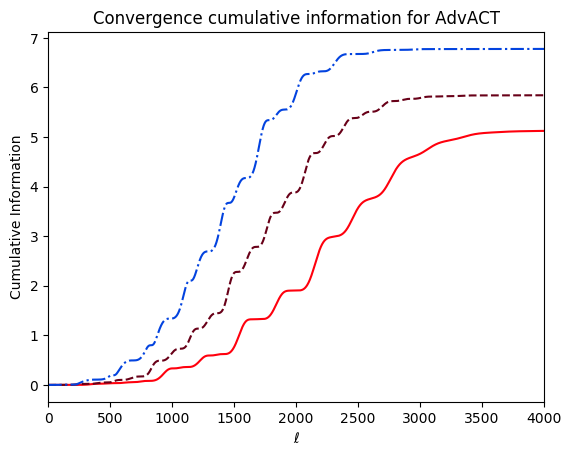}
  \end{minipage}
  \begin{minipage}{0.49\linewidth}
  \centering
  \includegraphics[width=\textwidth]{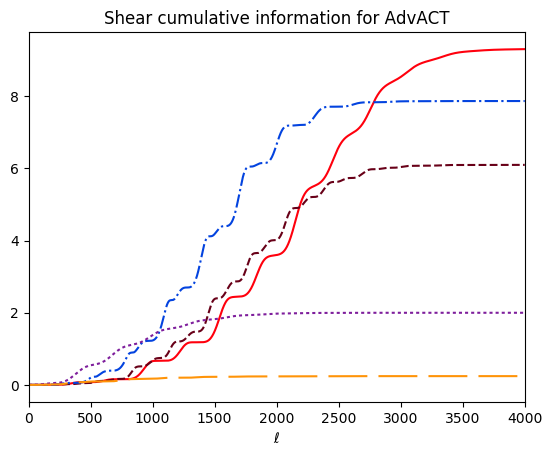}
  \end{minipage}
  
  \begin{minipage}{0.49\linewidth}
  \centering
  \includegraphics[width=\textwidth]{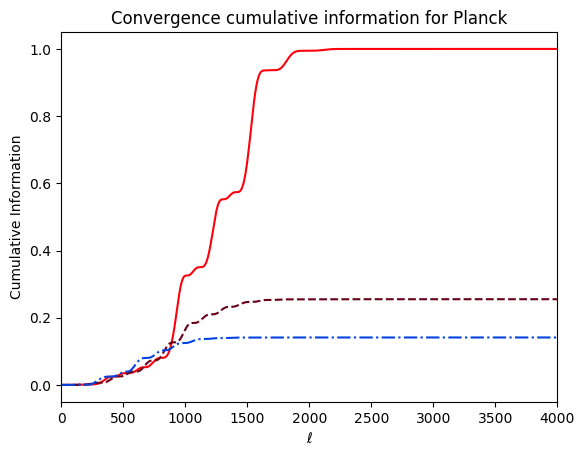}
  \end{minipage}
  \begin{minipage}{0.49\linewidth}
  \centering
  \includegraphics[width=\textwidth]{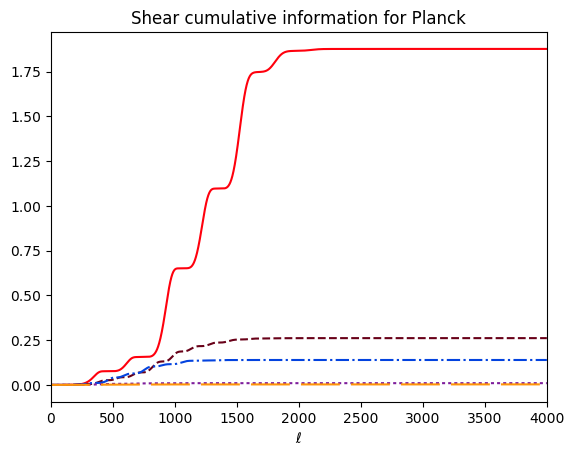}
  \end{minipage}
  \caption{Cumulative information for different temperature and polarization estimators, calculated using noise properties of  CMB-S4 (top) AdvACT (center) and Planck (bottom). Foregrounds are included in these plots.}
\label{fig:cumulative}
\end{figure}

The cumulative information for the Planck, AdvACT and CMB-S4 convergence and shear estimators is shown in figure \ref{fig:cumulative_no_FG}, which only includes a detector noise contribution, and figure \ref{fig:cumulative}, which additionally includes a foreground noise contribution.  All curves are normalized relative to the Planck convergence estimator. Note that both components of the shear have the same cumulative information and that correcting for the multiplicative bias (discussed in section \ref{sec:formfactor}) reduces the signal-to-noise for each lensing multipole $L,$ but does not change the relative contribution of each CMB multipole $\ell$ to the overall information. The structure in the cumulative information curves is related to the acoustic peak structure in the CMB spectra, and the increase in information for the convergence or shear estimators comes from scales where the CMB spectrum deviates from scale invariance or white noise, respectively. In all cases we observe that the bulk of information comes from large wavenumbers ($\ell \gtrsim 1000 $), which justifies the use of the squeezed triangle approximation in formulating these estimators. In the case of the $EB$ and $TB$ estimators, relatively more lensing information arises from larger scales compared to the $TT,$ $TE$ and $EE$ estimators because there is no cosmic variance in the $B$ mode variance term, and there is little additional information above $\ell \sim 1500,$ where the CMB $E$ mode spectrum falls sharply to zero.

In general we find that the shear estimators perform marginally better than the convergence estimators for the different map combinations and experiments, irrespective of foregrounds. For Planck the inclusion of foregrounds has little effect on the cumulative information because the foreground spectra only come to dominate the instrument noise on scales smaller than those where the bulk of the lensing information resides ($\ell \lesssim 2000$). It is evident that lensing estimators based on temperature maps dominate the signal-to-noise for Planck, as expected \citep{Planck2015Lensing}, due to the high noise level in the polarization maps. 

Both AdvACT and CMB-S4 have greater cumulative information than Planck, particularly on small scales ($\ell \gtrsim 2000$) due to their smaller beams and lower noise levels. For AdvACT, the $TT$ and $EE$ convergence and shear estimators contribute the bulk of the cumulative information, whereas for CMB-S4 the $EB$ shear estimator outperforms the other estimators and has significantly more cumulative information compared to AdvACT due to the lower instrumental polarized noise level, which is expected following our discussion in the introduction. The effect of foregrounds is larger for the AdvACT and CMB-S4 cumulative information curves than for Planck as the lower instrumental noise levels make the foregrounds relatively more pronounced and dominant over small scale temperature and polarization anisotropies that contain lensing information. Hereafter all results presented will incorporate foregrounds.

\section{Real space estimators as a limit of harmonic space estimators}
\label{sec:RS_from_HS}
The real space estimators presented above can be derived in the low $L$ limit of the harmonic space estimators traditionally used for CMB lensing reconstruction \citep{Seljak1999, Hu&Okamoto2002, Cooray2003}. The derivation presented below generalizes a similar calculation in ref. \cite{Bucher2012} for temperature estimators  to include polarization estimators. Harmonic space estimators make use of the fact that lensing induces correlations that are proportional to the lensing potential between previously uncorrelated CMB modes \citep{Seljak1996, Metcalf&Silk1997}
      \begin{equation}
      \left\langle X\left(\frac{\boldsymbol{L}}{2}-\boldsymbol{l}\right)Y\left(\frac{\boldsymbol{L}}{2}+\boldsymbol{l}\right) \right\rangle = f_{XY}(\boldsymbol{l},\boldsymbol{L})\,\,\phi(\boldsymbol{L}).
      \label{eq:f_XY}
      \end{equation}
We evaluate the lensing fields at $\frac{\boldsymbol{L}}{2}\pm\boldsymbol{\ell}$ so that we can use a two-sided difference when taking the low $L$ limit.
       
       The harmonic space quadratic estimator for the lensing potential is a weighted average of these off-diagonal modes \citep{Hu&Okamoto2002}:
      \begin{equation}
      \hat{\psi}_q^{XY}(\boldsymbol{L})= \frac{1}{N_q^{XY}(\boldsymbol{L})} \int \frac{d^2\boldsymbol{\ell}}{(2\pi)^2}
      \tilde{X}\left(\frac{\boldsymbol{L}}{2}-\boldsymbol{\ell}\right)\tilde{Y}^*\left(\frac{\boldsymbol{L}}{2}+\boldsymbol{\ell}\right) g_q^{XY}(\boldsymbol{\ell},\boldsymbol{L})\,,
      \label{eq:psi_main}
      \end{equation}
      where $N_q^{XY}(\boldsymbol{L})$ is a normalization factor to make the estimator unbiased
       and $g_q^{XY}(\boldsymbol{\ell},\boldsymbol{L})$ is a weight function that minimizes the variance.
  Taking the squeezed triangle limit, where $L \ll l$, and choosing the $x$-axis, and therefore the $\ell_x$-axis, to be aligned with the lensing wavevector $\boldsymbol{L}$, the harmonic space normalization and weight function can be related to the real space ones by  
      \begin{equation}
      N_q^{XY}=\frac{1}{4}L^4\left(N_{\hat{\kappa}_0}^{XY}+ N_{\hat{\gamma}_+}^{XY}\right)
      \label{eq:N_XY}
      \end{equation}
      and
      \begin{equation}
      g_q^{XY}=\frac{1}{2} L^2 \left(g_{\hat{\kappa}_0}^{XY} +  g_{\hat{\gamma}_+ }^{XY} \right),
      \label{eq:g_XY}
      \end{equation}
      for $XY=TT$, $EE$ and $TE$ where $N_{{\kappa}_0}^{XY}$,  $N_{{\gamma}_+ }^{XY}$, $g_{{\kappa}_0}^{XY} $ and $g_{{\gamma}_+ }^{XY} $ are given in table \ref{table:estimators}. Similarly, for the $EB$ and $TB$ combinations we have 
	\begin{equation}
      N_q^{YB}=\frac{1}{4}L^4\,N_{\hat{\gamma}_+}^{YB}
      \label{eq:N_YB}
      \end{equation}
      and
      \begin{equation}
      g_q^{YB}=\frac{1}{2} L^2 \,g_{\hat{\gamma}_+ }^{YB},
      \label{eq:g_YB}
      \end{equation}
where $N_{{\gamma}_+ }^{YB}$ and $g_{{\gamma}_+ }^{YB} $ are given in table \ref{table:estimators}.
Further details of the above calculations are given in appendix \ref{app:HS}. 

      The choice to align the $x$-axis with the lensing wavevector (by setting $\phi_L$ in figure \ref{fig:coords} to zero) corresponds to measuring shear $E$ and $B$ modes $\gamma_E$ and $\gamma_B$, where  $\gamma_E (\boldsymbol{L})$ corresponds to $\gamma_+(\boldsymbol{L})$ in a frame aligned with the lensing wavevector and $\gamma_B(\boldsymbol{L})$ corresponds to $\gamma_\times(\boldsymbol{L})$. For any other choice of axes, the shear $E$ and $B$ modes can be obtained by rotating $\gamma_+(\boldsymbol{L})+i\gamma_\times(\boldsymbol{L})$ into the basis aligned with the wavenumber $\boldsymbol{L}$ to obtain 
      \begin{equation}
      \gamma_E(\boldsymbol{L})+i\gamma_B(\boldsymbol{L})=e^{-2i\phi_L}\left[\gamma_+(\boldsymbol{L})+i\gamma_\times(\boldsymbol{L})\right].
      \end{equation}
The convergence and shear are related to the lensing potential in harmonic space by
\begin{equation}
\kappa_0(\boldsymbol{L})=\frac{1}{2} L^2 \psi(\boldsymbol{L})  \qquad
\gamma_+(\boldsymbol{L})=\frac{1}{2} (L_x^2-L_y^2)  \psi(\boldsymbol{L}) \qquad
\gamma_\times(\boldsymbol{L})=L_x L_y  \psi(\boldsymbol{L}),
\label{eq:pot_relations}
\end{equation}
where $\boldsymbol{L}=(L_x,L_y)$ is the lensing wavevector in the flat sky approximation. For shear $E$ and $B$ modes these relations become
\begin{equation}
\gamma_E(\boldsymbol{L})=\frac{1}{2} L^2 \psi(\boldsymbol{L}), \qquad
\gamma_B(\boldsymbol{L})=0, 
\label{eq:pot_relations_EB}
\end{equation}
which demonstrates that weak lensing produces only shear $E$ modes and no shear $B$ modes.
      
Having obtained the normalization and weight functions in the low $L$ limit, it is straightforward to show that the harmonic space lensing potential estimator is given in terms of the real-space lensing convergence and shear estimators by 
      \begin{equation}
     \frac{1}{2} L^2 \hat{\psi}_q^{XY}(\boldsymbol{L})\approx \frac{N^{XY}_{\hat{\kappa}_0}(\boldsymbol{L})}{N^{XY}_{\hat{\kappa}_0}(\boldsymbol{L})+N^{XY}_{\hat{\gamma}_E}(\boldsymbol{L})}\hat{\kappa}^{XY}_0(\boldsymbol{L})
+\frac{N^{XY}_{\hat{\gamma}_E}(\boldsymbol{L})}{N^{XY}_{\hat{\kappa}_0}(\boldsymbol{L})+N^{XY}_{\hat{\gamma}_E}(\boldsymbol{L})}\hat{\gamma}^{XY}_E(\boldsymbol{L})
      \end{equation}
      for $XY=TT$, $EE$ and $TE,$ where $\hat{\kappa}_0^{XY}$ and $\hat{\gamma}_E^{XY}$ are given in section \ref{sec:RS_implementation}. 
	The harmonic space estimator for the lensing potential in the low $L$ limit is therefore the inverse variance weighted combination of the real space convergence and shear $E$ mode estimators.
	Similarly for the $TB$ and $EB$ estimators we have
	   \begin{equation}
	\frac{1}{2}  L^2
	   \hat{\psi}_q^{YB}(\boldsymbol{L})\approx \hat{\gamma}_E^{YB},
	   \end{equation}
	   where $\hat{\gamma}_E^{YB}$ corresponds to $\hat\gamma^{YB}_+$ in our choice of coordinates.
The convergence estimator is absent for the $TB$ and $EB$ combinations as the large-scale convergence fields in our approximation do not generate $B$ modes.
 We have thus demonstrated that in the limit of large-scale lensing fields the harmonic space estimator for the lensing potential is a weighted sum of the real space convergence and shear estimators.

\section{Real space estimators from T,Q and U correlation functions}
\label{sec:corr_fn}

The estimators presented in section \ref{sec:RS_implementation}, based on $E$ and $B$ mode polarization maps, are the real-space analogues of the quadratic polarization lensing estimators in harmonic space \citep{Hu&Okamoto2002} in the squeezed triangle approximation. However, these estimators are not truly local because  $E$ and $B$ modes are defined nonlocally: they depend on the alignment of the polarization with nonlocal Fourier wavevectors \citep{KKS1997, Zaldarriaga&Seljak1997}. Methods of locally converting from maps of the $Q$ and $U$ Stokes parameters to $E$ and $B$ maps have been developed  \citep{Lewis2002, Bunn2003, Zhao2010}, but the most direct and intuitive way to reconstruct the lensing field in map space is to use $Q$ and $U$ rather than $E$ and $B$  polarization maps.

In this section, we present an alternative derivation that deals directly with CMB correlation functions in real space, resulting in lensing estimators that are applied to temperature and $Q$ and $U$ polarization maps. These estimators can be applied to observational CMB maps without requiring nonlocal treatments to construct the polarization $E$ and $B$ mode maps. Simulated maps with periodic boundary conditions can easily be used to create $E$ and $B$ mode maps, so we will show reconstructions from simulated CMB maps that make use of the $E$ and $B$ estimators defined above. However, for experimental data the estimators that act on $Q$ and $U$ maps will be the most straightforward to apply.

\subsection{Lensed correlation functions}
We consider the temperature and polarization correlation functions \citep{Lewis&Challinor2006}
\begin{align}
\xi_T(r)&=\langle T(\boldsymbol{x})T(\boldsymbol{x}-\boldsymbol{r}) \rangle
\nonumber
\\
\xi_p(r)&=\langle P_r(\boldsymbol{x})P_r^*(\boldsymbol{x}-\boldsymbol{r})\rangle =\langle P(\boldsymbol{x})P^*(\boldsymbol{x}-\boldsymbol{r})\rangle 
\nonumber
\\
\xi_m(r)&=\langle P_r(\boldsymbol{x})P_r(\boldsymbol{x}-\boldsymbol{r})\rangle =\langle e^{-4i\phi_r}P(\boldsymbol{x})P(\boldsymbol{x}-\boldsymbol{r})\rangle 
\nonumber
\\
\xi_c(r)&=\langle P_r(\boldsymbol{x})T(\boldsymbol{x}-\boldsymbol{r})\rangle =\langle e^{-2i\phi_r}P(\boldsymbol{x})T(\boldsymbol{x}-\boldsymbol{r})\rangle 
 \label{eq:corr_fns}
\end{align}
where $P(\boldsymbol{x})=Q(\boldsymbol{x})+iU(\boldsymbol{x})$, and  $P_r(\boldsymbol{x})=e^{-2i\phi_r}P(\boldsymbol{x})$ is the polarization field expressed in the physically relevant basis defined by $\boldsymbol{r}$, with $\phi_r$ the polar angle of $\boldsymbol{r}$ \citep{Lewis&Challinor2006}. Thus $Q_r$ describes polarization along the direction $\hat{r}$ ($Q_r$ positive) or perpendicular to $\hat{r}$ ($Q_r$ negative), and $U_r$ describes polarization at a $45^\circ$ angle to these directions. In ref. \citep{Lewis&Challinor2006} $\xi_p$, $\xi_m$ and $\xi_c$ are denoted as $\xi_+$, $\xi_-$ and $\xi_\times$ respectively but we use different notation to avoid confusion when writing the $\gamma_+$ and $\gamma_\times$ estimators for the different correlation functions.

The lensed correlation functions are given by
\begin{equation}
\tilde{\xi}(\boldsymbol{r})=\langle \tilde{X} (\boldsymbol{x})\tilde{Y}(\boldsymbol{x}-\boldsymbol{r})  \rangle,
\end{equation}
where $X$ and $Y$ are $T$, $P_r$ and $P_r^*$, and the lensed maps can be related to the unlensed temperature and polarization maps using 
\begin{equation}
\tilde{X}(\boldsymbol{x})=X(e^{\boldsymbol{\kappa}}\boldsymbol{x})\approx X(\boldsymbol{x}+\boldsymbol{\kappa}\boldsymbol{x})\approx X(\boldsymbol{x})+\boldsymbol{\nabla} X(\boldsymbol{x})\cdot \boldsymbol{\kappa} \boldsymbol{x}.
\end{equation}
The lensed correlation functions can therefore be expressed in terms of the unlensed correlation functions defined in eqs. (\ref{eq:corr_fns})  and the lensing fields that make up the deformation tensor, in a region in which $\kappa_0$, $\gamma_+$ and $\gamma_\times$ are constant and small, as 
\begin{equation}
\tilde{\xi}(\boldsymbol{r})=\xi(r)+\frac{d\xi}{d\ln(r)} \left(\kappa_0+\gamma_+ \cos(2\phi_r)+ \gamma_\times \sin(2\phi_r)\right),
\end{equation}	
where $\xi={\xi}_T, {\xi}_p, {\xi}_m$ or ${\xi}_c$. A full derivation of this result is given in appendix \ref{app:lensed_corr}.

\subsection{Estimators}
The lensing convergence and shear can be estimated using a weighted average of the lensed correlation function over different separations $\boldsymbol{r}$:
\begin{align}
\hat{\kappa_0}&=\int d^2\boldsymbol{r} \mathcal{K}_{\kappa_0}(r) \tilde{\xi}(\boldsymbol{r}) \nonumber \\
\hat{\gamma}_+&=\int d^2\boldsymbol{r} \mathcal{K}_{\gamma}(r) \cos(2\phi_r) \tilde{\xi}(\boldsymbol{r})=\int d^2\boldsymbol{r} \mathcal{K}_{\gamma_+}(\boldsymbol{r}) \tilde{\xi}(\boldsymbol{r}) \nonumber \\
\hat{\gamma}_\times&=\int d^2\boldsymbol{r}\mathcal{ K}_{\gamma}(r) \sin(2\phi_r) \tilde{\xi}(\boldsymbol{r})=\int d^2\boldsymbol{r} \mathcal{K}_{\gamma_\times}(\boldsymbol{r}) \tilde{\xi}(\boldsymbol{r}), \label{eq:corr_ests}
\end{align}
where $\mathcal{K}_{\gamma_+}(\boldsymbol{r}) \equiv \mathcal{K}_{\gamma}(r) \cos(2\phi_r)$, $\mathcal{K}_{\gamma_\times}(\boldsymbol{r}) \equiv \mathcal{K}_{\gamma}(r) \sin(2\phi_r)$ and $\tilde \xi(\boldsymbol{r})$ is one of the four lensed correlation functions from eq. (\ref{eq:corr_fns}). The lensing kernels, which are derived below, normalize the estimators and weight the integral according to which separations contribute most to the reconstruction.

\subsubsection*{Kernels for correlation function estimators}
The optimal lensing reconstruction kernels can be found by relating the estimators in eq. (\ref{eq:corr_ests}), and thus their kernels, to the real-space $T$, $E$ and $B$ estimators in section \ref{sec:estimators}. 
For example, the lensing convergence estimator based on the temperature correlation function $\xi_T$ is given by
\begin{align}
\hat{\kappa_0}^T&=\int d^2\boldsymbol{r} \mathcal{K}^T_{\kappa_0}(r) \tilde{\xi}_T(\boldsymbol{r}) =\int d^2\boldsymbol{r} \mathcal{K}^T_{\kappa_0}(r) \int d^2\boldsymbol{x} T(\boldsymbol{x})T(\boldsymbol{x}-\boldsymbol{r}) 
\nonumber 
\\
&=\int d^2\boldsymbol{x} T(\boldsymbol{x}) (\mathcal{K}^T_{\kappa_0} \circ T)(\boldsymbol{x}).  
\end{align}
This takes the same form as the estimator in eq. (\ref{eq:RSimplementation}) derived in section \ref{sec:estimators}, and the kernels are therefore the same: $\mathcal{K}^T_{\kappa_0} = K_{\kappa_0}^{TT}$ as given in eq. (\ref{eq:KapKernel}). The kernel peaks at small angular scales and drops off quickly, so the cleanest reconstruction is obtained by only keeping the central pixel in the integrand as before. The estimator can be translated to different points of the map, giving 
\begin{equation}
\hat{\kappa_0}^T(\boldsymbol{x})=T(\boldsymbol{x}) (\mathcal{K}^T_{\kappa_0} \circ T)(\boldsymbol{x}).
\end{equation}

Similarly, the lensing convergence estimator based on $\xi_p$ after translation to different pixels on the map is
\begin{equation}
\hat{\kappa_0}^p(\boldsymbol{x})=\tilde{Q}(\boldsymbol{x}) (\mathcal{K}^p_{\kappa_0}  \circ \tilde{Q})(\boldsymbol{x}) + \tilde{U}(\boldsymbol{x}) ( \mathcal{K}^p_{\kappa_0} \circ \tilde{U})(\boldsymbol{x}).
\end{equation}
The reconstruction kernel can be obtained by taking the expectation value of the Fourier transform of the estimator and comparing it to the expressions for the estimators in section \ref{sec:estimators}. This gives
\begin{equation}
\langle \hat{\kappa}_0^p(\boldsymbol{\ell}) \rangle
=\int \frac{d^2 \boldsymbol{\ell}^\prime}{(2\pi)^2} \mathcal{K}^p_{\kappa_0}(\boldsymbol{\ell}^\prime) \left(\tilde{C}^{QQ}_{\boldsymbol{\ell}^\prime} + \tilde{C}^{UU}_{\boldsymbol{\ell}^\prime} \right).
\end{equation}
We can rewrite this in term of $E$ and $B$ modes using \citep{Lewis&Challinor2006}
\begin{equation}
\tilde{C}^{QQ}_{\boldsymbol{\ell}} + \tilde{C}^{UU}_{\boldsymbol{\ell}}   = \int d^2 \boldsymbol{r} \xi_p(\boldsymbol{r}) e^{-i \boldsymbol{\ell} \cdot \boldsymbol{r}} 
=\tilde{C}^{EE}_{\boldsymbol{\ell}} +\tilde{C}^{BB}_{\boldsymbol{\ell}}  
\approx \tilde{C}^{EE}_{\boldsymbol{\ell}} ,
\end{equation}
where the last equality holds because in our first order approximation $\tilde{C}^{BB}(\boldsymbol{\ell})\approx0$.
The convergence estimator thus simplifies to 
\begin{equation}
\langle \hat{\kappa}_0^p(\boldsymbol{\ell}) \rangle=\int \frac{d^2 \boldsymbol{\ell}^\prime}{(2\pi)^2} \mathcal{K}^p_{\kappa_0}(\boldsymbol{\ell}^\prime) \tilde{C}^{EE}_{\boldsymbol{\ell}^\prime} ,
\end{equation}
which matches the expectation value of the $EE$ estimator in eq. (\ref{eq:kappaXY}). Thus $\mathcal{K}^p_{\kappa_0}(\boldsymbol{\ell})$ for the $\xi_p$ estimator is the same kernel as for the $EE$ estimator, given in eq. (\ref{eq:KapKernel}).

Similar comparisons for the lensing shear estimators based on ${\xi}_T$ and ${\xi}_p$, and for shear and convergence estimators based on ${\xi}_m$ and $\xi_c$ allow us to find their kernels in terms of those defined in section \ref{sec:estimators}. Table \ref{table:corr_estimators} gives the forms taken by the convergence and shear plus estimators respectively, as well as the expressions for the relevant kernels. 
The estimators for  $\gamma_\times$ are the same as those for $\gamma_+$ but with the factors of $\cos$ replaced by $\sin$ and vice versa.	
These estimators are provided for completeness, and for future application to experimental CMB maps. However, when we apply our real space estimators to simulated maps in section \ref{sec:reconstructions}, we make use of the estimators defined in section \ref{sec:RS_implementation} that act on $T$, $E$ and $B$ maps.

\begin{table}[!tbp]
\footnotesize	
\centering
\begin{tabular}{|c| c| c|}
\hline
Estimator & Form & Kernel \\ 

\hline

 $\hat{\kappa}_0^T$ & $\tilde{T}(\boldsymbol{x}) (\mathcal{K}^T_{\kappa_0} \circ \tilde{T})(\boldsymbol{x})$ & $\mathcal{K}^T_{\kappa_0}=\frac{1}{2 \pi N_{\kappa_0}^{TT}} \int_0^{\infty}  \ell d\ell J_0(\ell x) g_{\kappa_0}^{TT}(\ell) = K_{\kappa_0}^{TT}$ \\ 
 
  $\hat{\kappa}_0^p$ & $\tilde{Q}(\boldsymbol{x}) (\mathcal{K}^p_{\kappa_0} \circ \tilde{Q})(\boldsymbol{x})+\tilde{U}(\boldsymbol{x}) (\mathcal{K}^p_{\kappa_0} \circ \tilde{U})(\boldsymbol{x})$ & $\mathcal{K}^p_{\kappa_0}=\frac{1}{2 \pi N_{\kappa_0}^{EE}} \int_0^{\infty}  \ell d\ell J_0(\ell x) g_{\kappa_0}^{EE}(\ell)=K_{\kappa_0}^{EE} $ \\ 
 
  $\hat{\kappa}_0^m$ & $\tilde{Q}(\boldsymbol{x}) (\mathcal{K}^m_{\kappa_0} \circ \tilde{Q})(\boldsymbol{x})-\tilde{U}(\boldsymbol{x}) (\mathcal{K}^m_{\kappa_0} \circ \tilde{U})(\boldsymbol{x})$ & $\mathcal{K}^m_{\kappa_0}=\frac{1}{2 \pi N_{\kappa_0}^{EE}} \int_0^{\infty}  \ell d\ell J_0(\ell x) g_{\kappa_0}^{EE}(\ell)=K_{\kappa_0}^{EE} $ \\ 
  
  $\hat{\kappa}_0^c$ & $\tilde{Q}(\boldsymbol{x}) (\mathcal{K}^{c1}_{\kappa_0} \circ \tilde{T})(\boldsymbol{x})+\tilde{U}(\boldsymbol{x}) (\mathcal{K}^{c2}_{\kappa_0} \circ \tilde{T})(\boldsymbol{x})$ &       $\mathcal{K}^{c1}_{\kappa_0} =\left(\frac{1}{2 \pi N_{\kappa_0}^{TE}} \int_0^{\infty}  l d\ell J_0(\ell x) g_{\kappa_0}^{TE}(\ell)\right) \cos2\phi_r $ \\ 
  && $\mathcal{K}^{c2}_{\kappa_0} =\left(\frac{1}{2 \pi N_{\kappa_0}^{TE}} \int_0^{\infty}  \ell d\ell J_0(\ell x) g_{\kappa_0}^{TE}(\ell)\right) \sin2\phi_r $ \\ 
  
\hline
 $\hat{\gamma}_+^T$ & $\tilde{T}(\boldsymbol{x}) (\mathcal{K}^T_{\gamma_+} \circ \tilde{T})(\boldsymbol{x})$ &  $\mathcal{K}^T_{\gamma_+}=\left(\frac{1}{2 \pi N_{\gamma_+}^{TT}} \int_0^{\infty}  \ell d\ell J_2(\ell x) g_{\gamma_+}^{TT}(\ell )\right) \cos 2\phi_r $ \\ 

     $\hat{\gamma}_+^p$ & $\tilde{Q}(\boldsymbol{x}) (\mathcal{K}^p_{\gamma_+} \circ \tilde{Q})(\boldsymbol{x})+\tilde{U}(\boldsymbol{x}) (\mathcal{K}^p_{\gamma_+}  \circ \tilde{U})(\boldsymbol{x})$ & $\mathcal{K}^p_{\gamma_+} =\left(\frac{1}{2 \pi N_{\gamma_+}^{EE}} \int_0^{\infty}  \ell d\ell J_2(\ell x) g_{\gamma_+}^{EE}(\ell )\right) \cos 2\phi_r$ \\    

     $\hat{\gamma}_+^m$ & $\tilde{Q}(\boldsymbol{x}) (\mathcal{K}^m_{\gamma_+}  \circ \tilde{Q})(\boldsymbol{x})-\tilde{U}(\boldsymbol{x}) (\mathcal{K}^m_{\gamma_+}  \circ \tilde{U})(\boldsymbol{x})$ & $\mathcal{K}^m_{\gamma_+} =\left(\frac{1}{2 \pi } \int_0^{\infty}  \ell d\ell J_2(\ell x) \left(\frac{1}{\sqrt{2}} \frac{g_{\gamma_+}^{EE}(\ell )}{N_{\gamma_+}^{EE}}-\frac{1}{\sqrt{2}}\frac{g_{\gamma_+}^{EB}(\ell )}{N_{\gamma_+}^{EB}}\right)\right) \cos 2\phi_r + $ \\
     && $+ \left(\frac{1}{2 \pi } \int_0^{\infty}  \ell d\ell J_6(\ell x) \left(\frac{1}{\sqrt{2}} \frac{g_{\gamma_+}^{EE}(\ell )}{N_{\gamma_+}^{EE}}+\frac{1}{\sqrt{2}} \frac{g_{\gamma_+}^{EB}(\ell )}{N_{\gamma_+}^{EB}}\right)\right) \cos 6\phi_r$ \\ 

    $\hat{\gamma}_+^c$ & $\tilde{Q}(\boldsymbol{x}) (\mathcal{K}^{c1}_{\gamma_+}  \circ \tilde{T})(\boldsymbol{x})+\tilde{U}(\boldsymbol{x}) (\mathcal{K}^{c2}_{\gamma_+}  \circ \tilde{T})(\boldsymbol{x})$ &       $\mathcal{K}^{c1}_{\gamma_+} =\left(\frac{1}{4 \pi } \int_0^{\infty}  \ell d\ell J_0(\ell x) \left(\frac{1}{\sqrt{2}} \frac{g_{\gamma_+}^{TE}(\ell )}{N_{\gamma_+}^{TE}}-\frac{1}{\sqrt{2}}\frac{g_{\gamma_+}^{TB}(\ell )}{N_{\gamma_+}^{TB}}\right)\right) +$\\
    &&$+ \left(\frac{1}{4 \pi } \int_0^{\infty}  \ell d\ell J_4(\ell x) \left(\frac{1}{\sqrt{2}} \frac{g_{\gamma_+}^{TE}(\ell )}{N_{\gamma_+}^{TE}}+\frac{1}{\sqrt{2}}\frac{g_{\gamma_+}^{TB}(\ell )}{N_{\gamma_+}^{TB}}\right)\right)\cos4\phi_r $ \\ 
  && $\mathcal{K}^{c2}_{\gamma_+} =\left(\frac{1}{4 \pi } \int_0^{\infty}  \ell d\ell J_4(\ell x) \left(\frac{1}{\sqrt{2}} \frac{g_{\gamma_+}^{TE}(\ell )}{N_{\gamma_+}^{TE}}+\frac{1}{\sqrt{2}}\frac{g_{\gamma_+}^{TB}(\ell )}{N_{\gamma_+}^{TB}}\right)\right)\sin4\phi_r $ \\ 

\hline
\end{tabular}
\caption{Convergence and shear estimators from real space correlation functions. The estimators for  $\gamma_\times$ are the same as those for $\gamma_+$ but with the factors of $\cos$ replaced by $\sin$ and vice versa.}
\label{table:corr_estimators}
\end{table}

\section{Multiplicative bias}
\label{sec:formfactor}
The real space estimators are optimal only in the limit of small lensing angular wavenumber ${L}$, i.e., for lensing fields that vary over much larger scales than the CMB anisotropies. For lensing fields that vary over small angular scales, the amplitude of the reconstructed fields deviates from the input lensing field amplitude \citep{Bucher2012}. We can calculate this modulation of reconstructed power as a function of inverse angular scale by calculating how a plane wave lensing field of wavenumber $L$ is reconstructed by the estimators of section \ref{sec:RS_implementation}. 

If the convergence is given by the plane wave $\kappa_0(\boldsymbol{x})=A e^{i\boldsymbol{L}\cdot\boldsymbol{x}}$, or in Fourier space $\kappa_0(\boldsymbol{\ell})=(2\pi)^2 A \delta_D^2(\boldsymbol{\ell}-\boldsymbol{L})$, the lensed temperature in Fourier space in terms of the unlensed temperature and lensing field is
\begin{equation}
\tilde{T}(\boldsymbol{\ell}) \approx T(\boldsymbol{\ell}) - (\boldsymbol{\ell} \psi(\boldsymbol{\ell}))\circ (\boldsymbol{\ell} T(\boldsymbol{\ell})) =T(\boldsymbol{\ell})-\frac{2A}{L^2}\boldsymbol{L}\cdot(\boldsymbol{\ell}-\boldsymbol{L})T(\boldsymbol{\ell}-\boldsymbol{L}),
\end{equation}
where we Taylor expanded the lensed temperature before taking the Fourier transform. The convergence reconstructed using the real space estimator in eq. (\ref{eq:KappaEst}) and transformed into Fourier space (to allow us to study how the reconstruction depends on the angular wavenumber $L$) is given by
\begin{align}
\langle \hat{\kappa_0}(\boldsymbol{L}) \rangle &=\langle \tilde{T}(\boldsymbol{\ell})\circ\left(K(\boldsymbol{\ell})\tilde{T}(\boldsymbol{\ell})\right) \rangle 
\nonumber
\\  
&=\frac{2A}{L^2} \int {d^2\boldsymbol{\ell}^\prime} \boldsymbol{L}\cdot(\boldsymbol{\ell}^\prime-\boldsymbol{L})  C_{|\boldsymbol{\ell}^\prime-\boldsymbol{L}|}^{TT} \left(K(\boldsymbol{L}-\boldsymbol{\ell}^\prime)+ K(\boldsymbol{\ell}^\prime)\right).
\nonumber
\\  
&= \frac{2A}{L^2} \int {d^2\boldsymbol{\ell}^\prime} K(\boldsymbol{L}-\boldsymbol{\ell}^\prime) \left(\boldsymbol{L}\cdot\boldsymbol{\ell}^\prime	C_{\ell^\prime}^{TT} - \boldsymbol{L}\cdot(\boldsymbol{\ell}^\prime-\boldsymbol{L})  C_{|\boldsymbol{\ell}^\prime-\boldsymbol{L}|}^{TT}\right)
\end{align}
Dividing the reconstructed lensing field by the amplitude $(2\pi)^2 A$ of the input plane wave lensing field, we obtain the multiplicative bias,  which modulates the reconstructed lensing field amplitude as a function of wavenumber.  An equivalent calculation for $EE$, $TE$, $EB$ and $TB$, gives the multiplicative bias for the polarization estimators. The expressions for the multiplicative bias for the different temperature and polarization estimators are shown in table \ref{table:ff}. The multiplicative bias for the convergence and shear estimators can be obtained by setting the general kernel $K$ in the expressions in table \ref{table:ff} to their respective kernels from eqs. (\ref{eq:KapKernel2}) and (\ref{eq:ShearKernel2}).

\begin{table}[!tbp]
\centering
\begin{tabular}{|c |c|}
\hline
Estimator & Multiplicative Bias \\ 
\hline
 ${TT}$ & $\frac{2}{L^2} \int \frac{d^2\boldsymbol{\ell}^\prime}{(2\pi)^2} K(\boldsymbol{L}-\boldsymbol{\ell}^\prime) \left(\boldsymbol{L}\cdot\boldsymbol{\ell}^\prime	C_{\ell^\prime}^{TT} - \boldsymbol{L}\cdot(\boldsymbol{\ell}^\prime-\boldsymbol{L})  C_{|\boldsymbol{\ell}^\prime-\boldsymbol{L}|}^{TT}\right)$  \\ 
 
 ${EE}$ & $\frac{2}{L^2} \int \frac{d^2\boldsymbol{\ell}^\prime}{(2\pi)^2} \cos2(\phi_{\boldsymbol{\ell}^\prime-\boldsymbol{L}}-\phi_ {\boldsymbol{\ell}^\prime}) K(\boldsymbol{L}-\boldsymbol{\ell}^\prime) \left(\boldsymbol{L}\cdot\boldsymbol{\ell}^\prime C_{\ell^\prime}^{EE} - \boldsymbol{L}\cdot(\boldsymbol{\ell}^\prime-\boldsymbol{L}) C_{|\boldsymbol{\ell}^\prime-\boldsymbol{L}|}^{EE}\right)$ \\    
 
 ${TE}$& $\frac{2}{L^2} \int \frac{d^2\boldsymbol{\ell}^\prime}{(2\pi)^2} K(\boldsymbol{L}-\boldsymbol{\ell}^\prime) \left(\boldsymbol{L}\cdot\boldsymbol{\ell}^\prime  \cos2(\phi_{\boldsymbol{\ell}^\prime-\boldsymbol{L}}-\phi_ {\boldsymbol{\ell}^\prime})	C_{\ell^\prime}^{TE} - \boldsymbol{L}\cdot(\boldsymbol{\ell}^\prime-\boldsymbol{L})  C_{|\boldsymbol{\ell}^\prime-\boldsymbol{L}|}^{TE}\right)$  \\
    
 ${EB}$ & $\frac{2}{L^2} \int \frac{d^2\boldsymbol{\ell}^\prime}{(2\pi)^2} K(\boldsymbol{L}-\boldsymbol{\ell}^\prime) \boldsymbol{L}\cdot\boldsymbol{\ell}^\prime  \sin2(\phi_{\boldsymbol{\ell}^\prime-\boldsymbol{L}}-\phi_ {\boldsymbol{\ell}^\prime})	C_{\ell^\prime}^{EE}$\\
 
  ${TB}$ & $\frac{2}{L^2} \int \frac{d^2\boldsymbol{\ell}^\prime}{(2\pi)^2} K(\boldsymbol{L}-\boldsymbol{\ell}^\prime) \boldsymbol{L}\cdot\boldsymbol{\ell}^\prime  \sin2(\phi_{\boldsymbol{\ell}^\prime-\boldsymbol{L}}-\phi_ {\boldsymbol{\ell}^\prime})	C_{\ell^\prime}^{TE}$\\

\hline
\end{tabular}
\caption{Multiplicative bias for the lensing estimators, quantifying how the reconstructed lensing amplitude relates to the actual lensing amplitude as a function of wavenumber. The multiplicative bias expressions for the convergence and shear estimators are obtained by using the corresponding convergence and shear kernels of eqs. (\ref{eq:KapKernel}) and (\ref{eq:ShearKernel2}).}
\label{table:ff}
\end{table}

The multiplicative bias can be rewritten in map space in terms of the real space kernel and the temperature autocorrelation function as
\begin{align}
\frac{\langle\hat{\kappa_0}(\boldsymbol{L})\rangle}{\langle{\kappa_0}(\boldsymbol{L})\rangle}=&\frac{2}{L^2}\boldsymbol{L}\cdot  \int {d^2\boldsymbol{\theta}} \boldsymbol{\nabla} \left(C^{TT}(\boldsymbol{\theta}) K(\boldsymbol{\theta})\right) \delta(\boldsymbol{\theta}) e^{-i\boldsymbol{L}\cdot\boldsymbol{\theta}} + \nonumber \\ & +   \frac{2}{L^2} \int {d^2\boldsymbol{\theta}}\boldsymbol{\nabla} \cdot \left( \boldsymbol{\nabla} \left(C^{TT}(\boldsymbol{\theta})\right)  K(\boldsymbol{\theta}) \right) e^{-i\boldsymbol{L}\cdot\boldsymbol{\theta}} .
\end{align}
The first term vanishes because $\boldsymbol{\nabla} (C^{TT}(\boldsymbol{\theta}) K(\boldsymbol{\theta})) =0$ at $\boldsymbol{\theta}=0$. We use the second term to calculate the multiplicative bias in practice (as well as equivalent expressions for the polarization).

The multiplicative bias curves for temperature estimators applied to an experiment with Planck's sensitivity and resolution were presented in ref. \citep{Bucher2012}, using simulations of lensed temperature maps by plane wave lensing fields. Numerical errors due to the simulated maps having insufficient angular resolution to capture the large-scale reconstructed power resulted in the multiplicative bias being slightly overestimated in that analysis, and also resulted in different multiplicative bias curves for the two shear estimators. The analytical expression presented here allows us to compute the multiplicative bias with much higher accuracy, through a straightforward numerical integration.  

\begin{figure}[!thb]
\begin{minipage}{0.49\linewidth}
  \centering
  \includegraphics[width=\textwidth]{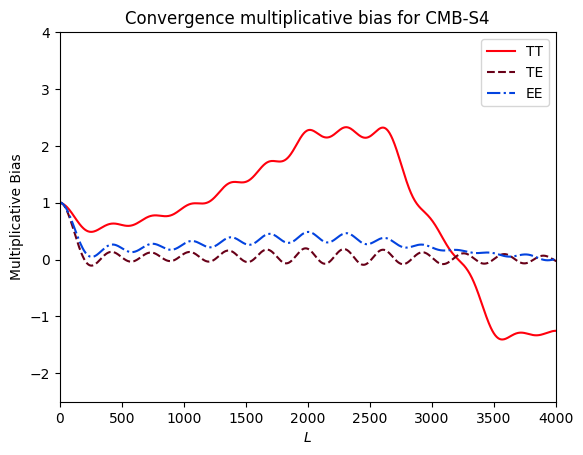}
  \end{minipage}
  \begin{minipage}{0.49\linewidth}
  \centering
  \includegraphics[width=\textwidth]{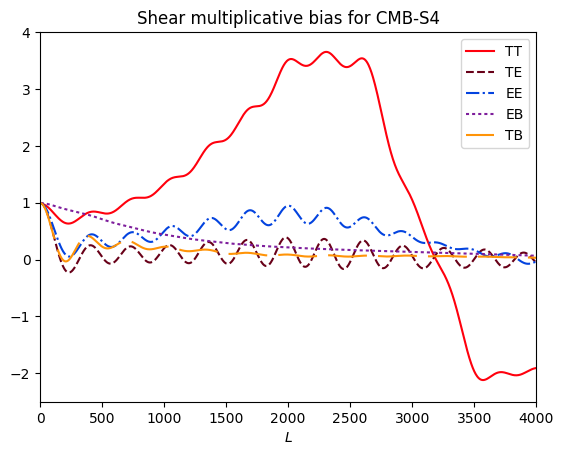}
  \end{minipage}

\begin{minipage}{0.49\linewidth}
  \centering
  \includegraphics[width=\textwidth]{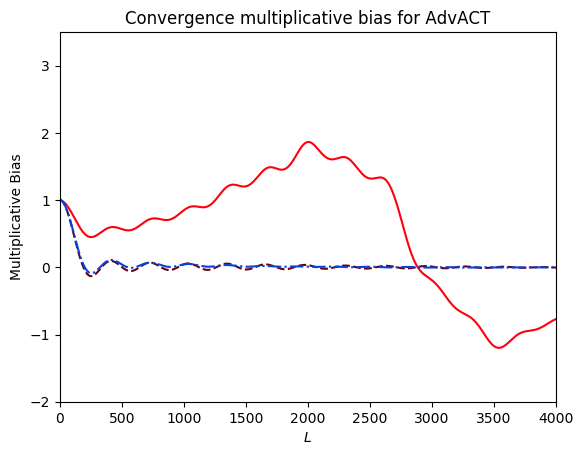}
  \end{minipage}
  \begin{minipage}{0.49\linewidth}
  \centering
  \includegraphics[width=\textwidth]{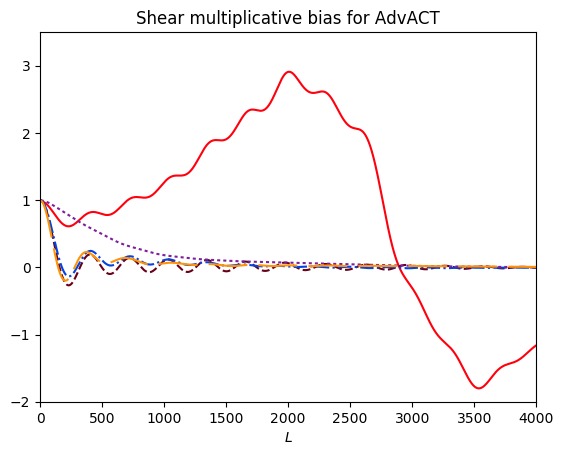}
  \end{minipage}
  
  \begin{minipage}{0.49\linewidth}
  \centering
  \includegraphics[width=\textwidth]{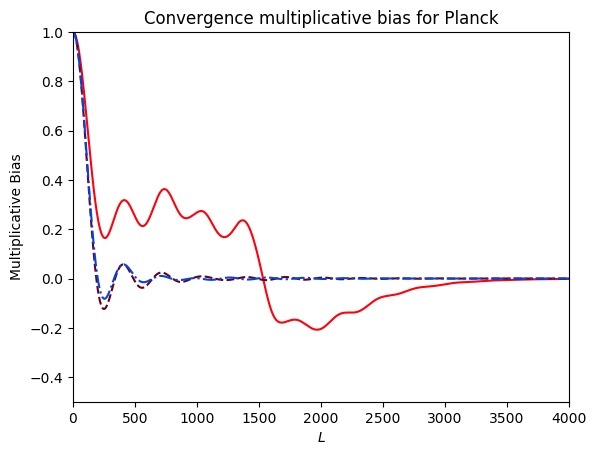}
  \end{minipage}
  \begin{minipage}{0.49\linewidth}
  \centering
  \includegraphics[width=\textwidth]{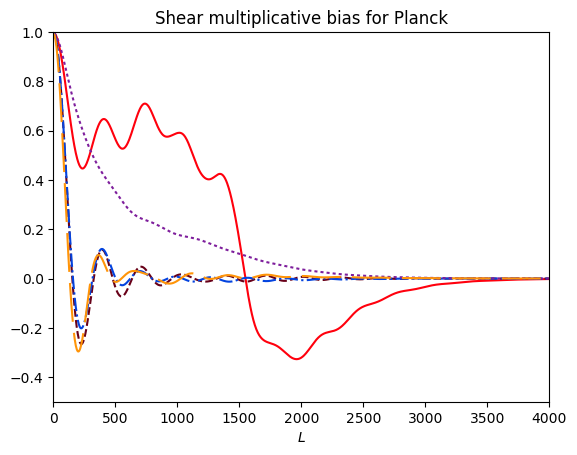}
  \end{minipage}
  \caption{Multiplicative bias for different temperature and polarization estimators, calculated using specifications from  CMB-S4 (top) AdvACT (center) and Planck (bottom). Foregrounds are included in the noise term when producting these plots.}
\label{fig:formfac}
\end{figure}

The CMB-S4, AdvACT, and Planck multiplicative bias curves are shown for the convergence (left column) and shear (right column) estimators in figure \ref{fig:formfac} (the shear plus and shear cross estimators have the same multiplicative bias). For large-scale lensing fields corresponding to small angular wave numbers $L\lesssim 100$, the lensing amplitude is accurately reconstructed because the squeezed triangle approximation is valid, while for smaller scale lensing fields (larger $L$) the amplitude of the reconstruction is biased and must be rescaled to accurately reconstruct the input lensing field. The temperature and $EB$ estimators have promising multiplicative bias curves, as the multiplicative bias falls off much less steeply with $L$ than for the other estimators. The multiplicative bias curves for the different experiments are similar in shape, with the larger beam and detector noise of Planck smearing out the lensing effect, leading to a smaller reconstructed amplitude than AdvACT and CMB-S4. 

The variation of the multiplicative bias with inverse angular scale can be understood in terms of the effect that the real-space kernel has on the lensing reconstruction. Considering for now the convergence amplitude reconstructed from CMB temperature ($TT$) maps, there is a drop in the multiplicative bias on large scales ($L < 200$) with increasing $L$. This is a consequence of the assumption that the convergence is constant over the angular range on which the lensing field distorts the CMB. The acoustic scale of $\sim$ 1 degree sets the coherence scale. 
If the convergence field decreases on this scale, as is the case for nonzero lensing wavenumbers, then the average convergence amplitude that is inferred in the lensing recontruction will be smaller than the true convergence amplitude.

On smaller scales ($L > 200$) the multiplicative bias for the convergence increases again, and exceeds unity in the case of AdvACT and CMB-S4 at $L \approx 1200$ and $L \approx 1000$, respectively. These lensing scales are smaller than the CMB coherence scale and as $L$ increases the reconstruction is increasingly dominated by squeezed triangle configurations where the lensed CMB and lensing wavenumbers are large and the unlensed CMB lensing wavenumber is small. This is well known: the small-scale lensing effect originates from the coupling of the small scale lensing field to the large-scale CMB gradient to produce a small-scale CMB anisotropy. This highlights that the weighting of small-scale CMB configurations should be determined by the relative angle of the CMB gradient to the lensing wavevector, with orthogonal CMB ($\boldsymbol{\ell}$) and lensing ($\mathbf{L}$) angular wavevectors producing no lensing effect. The isotropic real-space kernel, however, weights all relative orientations equally, thereby overestimating the true lensing amplitude.

On even smaller scales ($L > 1500$ for Planck and $L > 3000$ for AdvACT and CMB-S4) the multiplicative bias becomes negative. The fact that the form factor crosses zero, becoming negative and oscillating a few times, may at first sight seem paradoxical. But this behavior is not so different from that of certain low-pass filters, and in a certain sense our estimator may be viewed as a low-pass filter. The Fourier transform of a top-hat profile low-pass filter in two dimensions, for example, has the functional form $J_1(x)/x,$ thus exhibiting similar qualitative behavior.
On the other hand, the $EB$ multiplicative bias is always positive (being a convolution of two positive quantities) that peaks at large scales and smoothly decreases to zero.

The multiplicative bias curves in figure \ref{fig:formfac} model how the reconstructed lensing amplitude varies with inverse angular scale, allowing us to correct for this modulation of reconstructed power and obtain unbiased reconstructions of the convergence and shear power spectra, at the expense of an increase in the variance of the reconstruction.

\section{Lensing reconstruction on simulated CMB maps}
\label{sec:reconstructions}

We now test the real space estimators presented in Section \ref{sec:RS_implementation} on simulated CMB maps. CMB temperature and polarization maps and lensing maps are simulated using the spectra produced by CAMB\footnote{\url{http://camb.info/}} \citep{Lewis1999}. The simulated lensing maps are used to shift the unlensed CMB maps by the deflection angle to obtain lensed CMB maps in a $20^\circ$ by $20^\circ$ region of the sky with pixels of width 0.6 arcminutes, making use of a high resolution unlensed map (with a pixel scale of 0.3 arcminutes) to accurately simulate the deflection. Experimental beam smearing effects and experimental noise and foregrounds are included to obtain the final simulated lensed maps to which the real space estimators of eqs. (\ref{eq:KappaEst}) and (\ref{eq:GammaEst})  are applied.

\begin{figure}[!bht]
\begin{minipage}{0.49\linewidth}
  \centering
  \includegraphics[width=\textwidth]{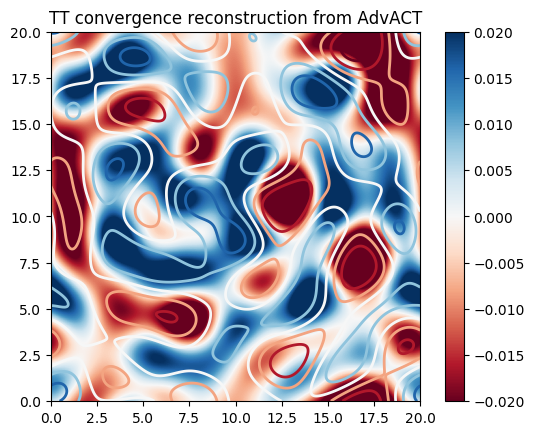}
  \end{minipage}
  \begin{minipage}{0.49\linewidth}
  \centering
  \includegraphics[width=\textwidth]{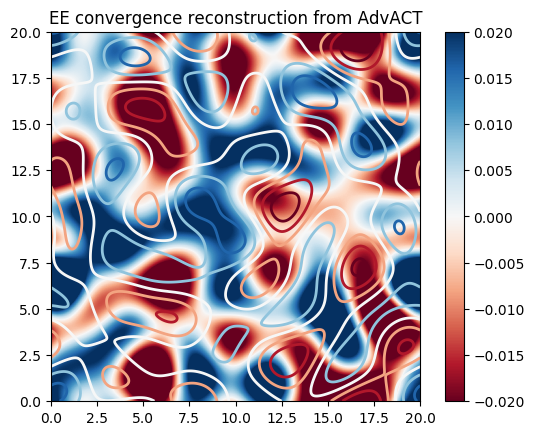}
  \end{minipage}
  
  \begin{minipage}{0.49\linewidth}
  \centering
  \includegraphics[width=\textwidth]{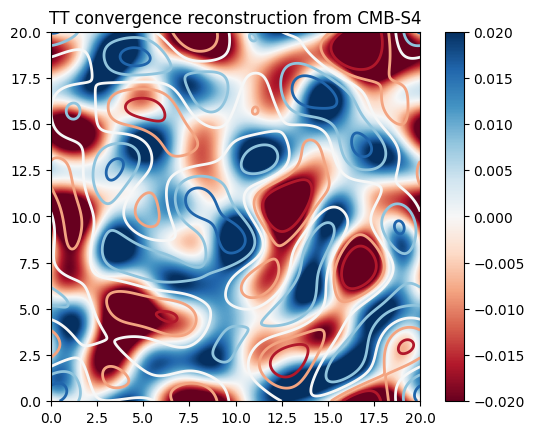}
  \end{minipage}
  \begin{minipage}{0.49\linewidth}
  \centering
  \includegraphics[width=\textwidth]{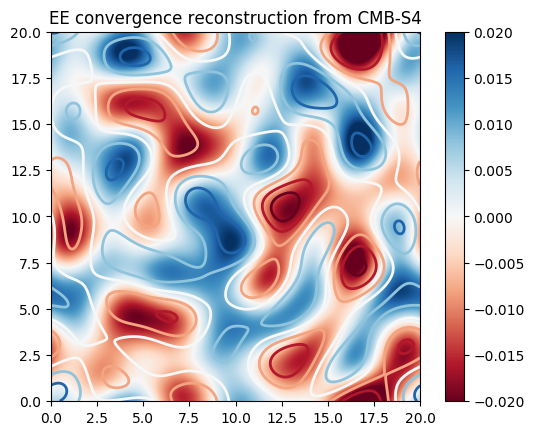}
  \end{minipage}

\caption{The real space convergence estimator applied to $20^\circ$ by $20^\circ$ lensed temperature maps (left) and E mode polarization maps (right). AdvACT noise (top row) and CMB-S4 noise (bottom row) has been added to the CMB maps and used in the reconstruction filter. The color map shows the reconstructed field while the contours show the actual lensing field used. The maps have been smoothed, keeping only Fourier modes with $15<\ell<100$.  }
\label{fig:rec_conv}
\end{figure}

\begin{figure}[!bht]
\begin{minipage}{0.49\linewidth}
  \centering
  \includegraphics[width=\textwidth]{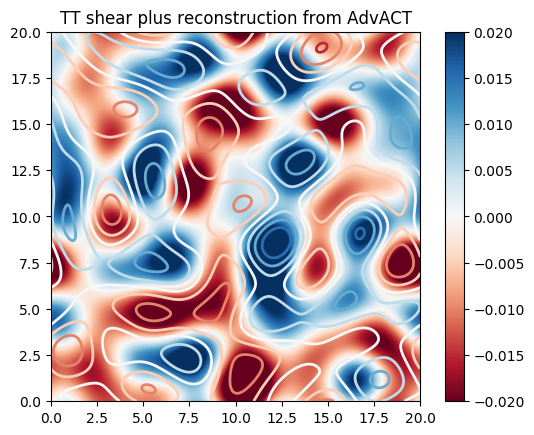}
  \end{minipage}
  \begin{minipage}{0.49\linewidth}
  \centering
  \includegraphics[width=\textwidth]{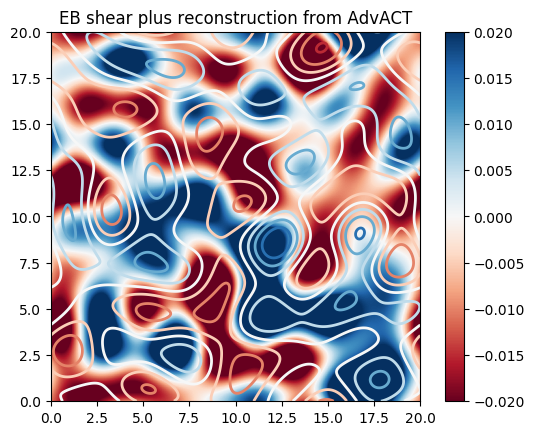}
  \end{minipage}
  
  \begin{minipage}{0.49\linewidth}
  \centering
  \includegraphics[width=\textwidth]{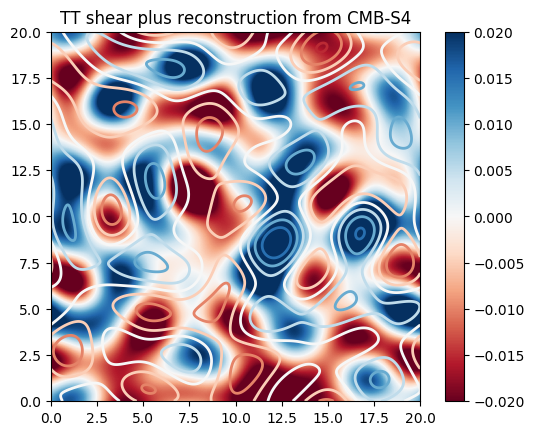}
  \end{minipage}
  \begin{minipage}{0.49\linewidth}
  \centering
  \includegraphics[width=\textwidth]{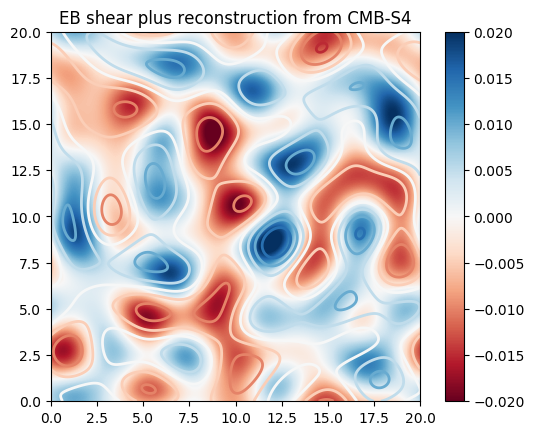}
  \end{minipage}

\caption{The real space shear plus estimator applied to $20^\circ$ by $20^\circ$ lensed temperature maps (left) and E and B mode polarization maps (right). AdvACT noise (top row) and CMB-S4 noise (bottom row) has been added to the CMB maps and used in the reconstruction filter.  The color map shows the reconstructed field while the contours show the actual lensing field used. The maps have been smoothed, keeping only Fourier modes with $15<\ell<100$. }
\label{fig:rec_shear}
\end{figure}

We compare the reconstructed lensing fields to the input lensing fields in map space, as this emphasizes the real space nature of the estimators and allows us to visually assess the accuracy of the reconstruction. It is harder to achieve visual agreement in map space than in the power spectrum as high signal to noise is needed for individual modes. The maps have been smoothed, keeping only Fourier modes with $15<\ell<100$ to show the modes that are not severely compromised by the multiplicative bias and to make the correspondence between the input and output maps clearer.  Map reconstructions are also useful for cross-correlation applications. The reconstructed convergence is shown in figure \ref{fig:rec_conv} for different combinations of temperature and E mode maps for AdvACT and CMB-S4 noise specifications, and the reconstructed shear for the $TT$ and $EB$ estimators is shown in figure \ref{fig:rec_shear}. The color map shows the reconstructed field while the contours show the actual lensing field used. The higher noise and poorer resolution of the Planck experiment mean that the map space reconstructions are dominated by noise on all scales, which is why they are not shown here.

The lensing reconstructions are affected by the instrumental noise of the experiment. For AdvACT noise, shown in the top rows of figures \ref{fig:rec_conv} and \ref{fig:rec_shear},  the reconstructed convergence is clearly correlated with the input convergence, but it is a noisy reconstruction, especially for the polarization estimators. The shear reconstructed from $E$ and $B$ maps is especially noisy, because the instrumental noise level for AdvACT means that maps of the lensed CMB $B$ modes are dominated by noise. For the upcoming CMB-S4 experiment, the noise in the reconstruction is significantly reduced and the polarization estimators give cleaner reconstructions than the CMB temperature. This is because the improved telescope sensitivity and resolution results in cleaner CMB maps, and the small-scale CMB modes that are most affected by lensing become available, giving access to more squeezed triangle configurations with well-measured CMB modes and allowing us to reconstruct the lensing effect with much higher signal to noise. Foregrounds dominate the CMB temperature on small scales \citep{Dunkley2013} and this is taken into account in the lensing kernel before applying these estimators to the simulated data. 

In the case of estimators that depend on $B$ modes, the spurious reconstruction signal from the unlensed fields is approximately zero, as primordial $B$ modes are negligibly small for our purposes. These estimators therefore do not have additional noise from cosmic variance, and so the $EB$ estimator gives the highest signal to noise reconstruction for futuristic experiments such as the CMB-S4 experiment \citep{Hu&Okamoto2002}. This is clear in figure \ref{fig:rec_shear}, which shows that the shear reconstruction from the CMB-S4 $EB$ estimator gives a very accurate, low-noise reconstruction on the scales shown.

\section{Discussion}
\label{sec:discussion}
The lensing reconstruction techniques described in this paper can be applied to temperature and polarization maps from current and future CMB experiments. While most methods of reconstructing the gravitational lensing potential make use of harmonic space lensing estimators, the real space estimators developed here are implemented locally in map space and could prove advantageous for CMB observations, which have point source excisions and non-uniform sky coverage, making it difficult to accurately obtain harmonic space quantities. The real space estimators developed in this paper also allow us to separately reconstruct the lensing convergence and the two components of the shear, which provides a useful consistency check of the method as these quantities are related via the lensing potential. The self-consistency of the convergence and shear maps reconstructed from CMB maps also provides a test of systematic errors in lensing reconstruction, particularly on large angular scales where the real space estimators perform competitively with their harmonic space counterpart. In a future paper we plan to apply these estimators to publicly available CMB maps.

As the sensitivity and resolution of CMB experiments continue to improve, the least noisy CMB lensing reconstructions will come from polarization \citep{Hu&Okamoto2002}.  
 The polarization estimators presented here will therefore prove useful in supplementing, and for future experiments improving on, the reconstruction obtained from the real space temperature estimator developed in ref. \citep{Bucher2012}. The temperature and polarization lensing reconstructions can then be combined to produce a minimum variance lensing map.

The main limitation of the lensing estimators presented in this paper is that the reconstructed lensing field is inaccurate for input lensing fields that vary on smaller angular scales ($L>100$), resulting in a multiplicative bias to the lensing amplitude. We showed that the multiplicative bias can be accurately calculated as a function of angular wavenumber, and thus corrected for, though at the expense of an increase in the noise of the reconstruction. However, it would be preferable to develop a real space estimator that accurately reconstructs the lensing field on all scales such that the multiplicative bias is one for all wavenumbers. This is possible by utilising an anisotropic lensing reconstruction kernel \citep{Carvalho2010}. Alternatively, a bilinear convolution scheme that appropriately weights the relative orientation of lensed CMB wavenumbers would account for a lensing field that varies on smaller angular scales. With such an estimator it becomes viable to consider CMB lensing power spectrum reconstruction without having to correct for a multiplicative bias term. In a future paper we aim to explore this extension to the real space lensing estimator considered here.


\section*{Acknowledgements}
HP acknowledges support from the SKA South Africa Project. KM acknowledges support from the National Research Foundation of South Africa (grant number 93565). MB acknowledges support from the National Institute of Theoretical Physics visitor program and a travel grant from the National Research Foundation of South Africa (grant number 104301).

\appendix
\section{Real space estimators from harmonic space estimators}
\label{app:HS}

In this appendix we derive the expressions given in eqs. (\ref{eq:N_XY}) - (\ref{eq:g_YB}), which relate the quadratic lensing estimator normalization, $N_q^{XY},$ and weight, $g_q^{XY},$ in the squeezed triangle limit, to the corresponding normalizations and weights of the convergence and shear estimators. The key quantity to evaluate is $f_{XY}(\boldsymbol{\ell}, \boldsymbol{L}),$ defined in eq. (\ref{eq:f_XY}), which is given by 
\begin{eqnarray}
  f_{XY}(\boldsymbol{\ell}, \boldsymbol{L}) &=&\boldsymbol{L}\cdot \left(\frac{\boldsymbol{L}}{2}-\boldsymbol{\ell}\right)  C^{XY}_{|\frac{\boldsymbol{L}}{2}-\boldsymbol{\ell}|} + \boldsymbol{L}\cdot \left(\frac{\boldsymbol{L}}{2}+\boldsymbol{\ell}\right)  C^{XY}_{|\frac{\boldsymbol{L}}{2}+\boldsymbol{\ell}|}  \qquad \mbox{for \,} XY = TT, EE, TE \nonumber \\
f_{YB}(\boldsymbol{\ell}, \boldsymbol{L})&=&\boldsymbol{L}\cdot \left(\frac{\boldsymbol{L}}{2}-\boldsymbol{\ell}\right)  C^{YE}_{|\frac{\boldsymbol{L}}{2}-\boldsymbol{\ell}|} \sin{(2 \phi_{\frac{\boldsymbol{L}}{2}+\boldsymbol{\ell},\frac{\boldsymbol{L}}{2}-\boldsymbol{\ell}})} \quad \qquad\qquad  \mbox{for \,}  YB = TB, EB
  \end{eqnarray}
We can expand $f_{XY}$ and $f_{YB}$ in the squeezed triangle limit ($L \ll \ell$) as follows
\begin{align}
  f_{XY}(\boldsymbol{\ell}, \boldsymbol{L})&=\boldsymbol{L}\cdot \left(\frac{\boldsymbol{L}}{2}-\boldsymbol{\ell}\right)  C^{XY}_{|\frac{\boldsymbol{L}}{2}-\boldsymbol{\ell}|} + \boldsymbol{L}\cdot \left(\frac{\boldsymbol{L}}{2}+\boldsymbol{\ell}\right)  C^{XY}_{|\frac{\boldsymbol{L}}{2}+\boldsymbol{\ell}|}\nonumber \\
  &\approx \frac{1}{2}{L^2}\left(\frac{1}{l^2}\frac{d(\ell^2C^{XY}_\ell)}{d\ln(\ell)}+ \left(\frac{2(\boldsymbol{\ell}\cdot\boldsymbol{L})^2}{L^2l^2}-1\right)\frac{dC^{XY}_\ell}{d\ln(\ell)}\right)\nonumber\\
  &\approx \frac{1}{2}{L^2}\left(\frac{1}{l^2}\frac{d(\ell^2C^{XY}_\ell)}{d\ln(\ell)}+ \cos(2\phi_{\ell L})\frac{dC^{XY}_\ell}{d\ln(\ell)}\right),
  \end{align}
while
\begin{align}
   f_{YB}(\boldsymbol{\ell},\boldsymbol{L})&\approx \left(\frac{L^2}{2}-\boldsymbol{\ell}\cdot\boldsymbol{L}\right)  C^{YE}_{\ell}
   \sin{(2 \phi_{\frac{\boldsymbol{L}}{2}+\boldsymbol{\ell},\frac{\boldsymbol{L}}{2}-\boldsymbol{\ell}})}\nonumber\\
   &\approx -\boldsymbol{\ell}\cdot\boldsymbol{L}\, C^{YE}_\ell \sin{(2 \phi_{\frac{\boldsymbol{L}}{2}+\boldsymbol{\ell},\frac{\boldsymbol{L}}{2}-\boldsymbol{\ell}})}\nonumber\\
   &\approx - 2 L^2 \cos(\phi_{\ell L})\sin(\phi_{\ell L}) C^{YE}_\ell \nonumber\\
   &\approx L^2 \sin(2\phi_{\ell L}) C^{YE}_\ell, 
\end{align}
where $\phi_{\ell L}$ is the angle between $\boldsymbol{\ell}$ and $\boldsymbol{L}$.  
It follows that the corresponding quadratic estimator weight functions,  
\begin{eqnarray}
  g_q^{XY}(\boldsymbol{\ell},\boldsymbol{L})&=&\frac{f_{XY}(\boldsymbol{\ell}, \boldsymbol{L})} {{\cal C}\,^{XY}_{|\frac{\boldsymbol{L}}{2}-\boldsymbol{\ell}|} \,  {\cal C}\,^{XY}_{|\frac{\boldsymbol{L}}{2}+\boldsymbol{\ell}|}} \qquad \qquad  \qquad \qquad\qquad \qquad\qquad \mbox{for \,} XY = TT, EE, \nonumber \\
  g_q^{TE}(\boldsymbol{\ell},\boldsymbol{L})&=&\frac{f_{TE}(\boldsymbol{\ell}, \boldsymbol{L}) \left({\cal C}\,^{TT}_{|\frac{\boldsymbol{L}}{2}-\boldsymbol{\ell}|} {\cal C}\,^{EE}_{|\frac{\boldsymbol{L}}{2}+\boldsymbol{\ell}|}-{\cal C}\,^{TE}_{|\frac{\boldsymbol{L}}{2}-\boldsymbol{\ell}|} {\cal C}\,^{TE}_{|\frac{\boldsymbol{L}}{2}+\boldsymbol{\ell}|}\right)}{{\cal C}\,^{TT}_{|\frac{\boldsymbol{L}}{2}-\boldsymbol{\ell}|} {\cal C}\,^{EE}_{|\frac{\boldsymbol{L}}{2}+\boldsymbol{\ell}|} {\cal C}\,^{EE}_{|\frac{\boldsymbol{L}}{2}-\boldsymbol{\ell}|} {\cal C}\,^{TT}_{|\frac{\boldsymbol{L}}{2}+\boldsymbol{\ell}|}-\left({\cal C}\,^{TE}_{|\frac{\boldsymbol{L}}{2}-\boldsymbol{\ell}|} \, {\cal C}\,^{TE}_{|\frac{\boldsymbol{L}}{2}+\boldsymbol{\ell}|}\right)^2},\nonumber \\
g_q^{YB}(\boldsymbol{\ell},\boldsymbol{L})&=&\frac{f_{YB}(\boldsymbol{\ell}, \boldsymbol{L})}{{\cal C}\,^{YY}_{|\frac{\boldsymbol{L}}{2}-\boldsymbol{\ell}|} \,\,n\,^{BB}(|\frac{\boldsymbol{L}}{2}+\boldsymbol{\ell}|)} \quad \qquad \qquad \qquad \qquad \qquad  \mbox{for \,} YB = TB, EB,
\end{eqnarray}
in which we define ${\cal C}\,^{XY}_\ell = C\,^{XY}_\ell+n\,^{XY}(\ell),$ can be simplified, using  
$C\,^{XY}_{|\frac{\boldsymbol{L}}{2}\pm\boldsymbol{\ell}|}+n^{XY}(|\frac{\boldsymbol{L}}{2}\pm\boldsymbol{\ell}|) \approx C\,^{XY}_\ell+n^{XY}(\ell),$ as follows
\begin{eqnarray}
  g_q^{XY}(\boldsymbol{\ell},\boldsymbol{L})&\approx & \frac{\frac{1}{2}{L^2}C^{XY}_\ell\left[\frac{d\ln(\ell^2C^{XY}_\ell)}{d\ln(\ell)}+ \cos(2\phi_{\ell L})\frac{d\ln(C^{XY}_\ell)}{d\ln(\ell)}\right]}{\left(C\,^{XY}_\ell+n^{XY}(\ell)\right)^2} \qquad \mbox{for \,} XY = TT, EE, \nonumber \\
  g_q^{TE}(\boldsymbol{\ell},\boldsymbol{L}) &\approx & \frac{\frac{1}{2}{L^2} \left[ \left(\frac{d\ell^2C^{TE}_\ell}{d\ln(\ell)}\right)+ \cos(2\phi_{\ell L})\frac{dC^{TE}_\ell}{d\ln(\ell)}\right]}{\left(C\,^{TT}_\ell+n^{TT}(\ell)\right)\left(C\,^{EE}_\ell+n^{EE}(\ell)\right)+\left(C\,^{TE}_\ell+n^{TE}(\ell)\right)^2}, \nonumber \\
g_q^{YB}(\boldsymbol{\ell},\boldsymbol{L})&\approx & \frac{L^2 \sin(2\phi_{\ell L}) C^{YE}_\ell}{\left(C\,^{YY}_\ell+n^{YY}(\ell)\right)\,n^{BB}(\ell) } \qquad \qquad \qquad \qquad \qquad  \mbox{for \,} YB = TB, EB. 
  \end{eqnarray}

The above expressions for $f_{XY}(\boldsymbol{\ell}, \boldsymbol{L}) $ and $g_q^{XY}(\boldsymbol{\ell}, \boldsymbol{L}) $ can be used to evaluate the quadratic estimator normalization 
\begin{equation}
  N_q^{XY}(\boldsymbol{L})=\int \frac{d^2\boldsymbol{\ell}}{(2\pi)^2} f_{XY}(\boldsymbol{\ell}, \boldsymbol{L}) g_q^{XY}(\boldsymbol{\ell}, \boldsymbol{L})
  \label{eq:LensingNoise}
  \end{equation}
in the squeezed triangle limit. 
If the $x$-axis and therefore our $l_x$-axis is aligned with the lensing wavevector $\boldsymbol{L}$ then $\phi_{\ell L}=\phi_{\ell},$ and it is straightforward to show that 
  \begin{eqnarray}
 	  g_q^{XY}&\approx&\frac{1}{2} L^2 \left(g_{\hat{\kappa}_0}^{XY} +  g_{\hat{\gamma}_+ }^{XY} \right) \quad \mbox{and} \nonumber \\
	 N_q^{XY}&\approx&\frac{1}{4}L^4(N_{\hat{\kappa}_0}^{XY}+ N_{\hat{\gamma}_+}^{XY}), 
  \end{eqnarray}
  where $g_{{\kappa}_0}^{XY}$,  $g_{{\gamma}_+ }^{XY}$, $N_{{\kappa}_0}^{XY} $ and $N_{{\gamma}_+ }^{XY} $ are given in table \ref{table:estimators} and $g_{{\kappa}_0}^{XY}$ is zero for $XY=EB$ and $TB$ as there are no convergence estimators for these combinations. 
\section{Lensed temperature correlation function}
\label{app:lensed_corr}
The lensed CMB temperature map $\tilde{T}(\boldsymbol{x})$ is related to the unlensed map ${T}(\boldsymbol{x})$ by
\begin{equation}
\tilde{T}(\boldsymbol{x})=T(e^{\boldsymbol{\kappa}}\boldsymbol{x})\approx T(\boldsymbol{x}+\boldsymbol{\kappa}\boldsymbol{x})\approx T(\boldsymbol{x})+\boldsymbol{\nabla} T(\boldsymbol{x})\cdot \boldsymbol{\kappa} \boldsymbol{x},
\end{equation}
where $\boldsymbol{\kappa}$ is the deformation tensor defined in eq. (\ref{eq:deformation_tensor}), provided that the lensing effect is small. Thus the lensed and unlensed temperature correlation functions are related by
\small
\begin{align}
\tilde{\xi}_{T}(\boldsymbol{r})&=\langle\tilde{T}(\boldsymbol{x}) \tilde{T}(\boldsymbol{x}+\boldsymbol{r})\rangle_{\boldsymbol{x}} \nonumber \\
&={\xi}_{T}(\boldsymbol{r})+\langle(\boldsymbol{\nabla} T(\boldsymbol{x}+\boldsymbol{r}) \cdot \kappa (\boldsymbol{x}+\boldsymbol{r})) {T}(\boldsymbol{x})\rangle_{\boldsymbol{x}} + \langle(\boldsymbol{\nabla} T(\boldsymbol{x}) \cdot \kappa \boldsymbol{x}) {T}(\boldsymbol{x}+\boldsymbol{r})\rangle_{\boldsymbol{x}} \nonumber \\
&={\xi}_{T}(\boldsymbol{r})+\langle(\boldsymbol{\nabla} T(\boldsymbol{x}+\boldsymbol{r}) \cdot \kappa \boldsymbol{x}) {T}(\boldsymbol{x})\rangle_{\boldsymbol{x}} +\langle(\boldsymbol{\nabla} T(\boldsymbol{x}+\boldsymbol{r}) \cdot \kappa \boldsymbol{r}) {T}(\boldsymbol{x})\rangle_{\boldsymbol{x}} + \langle(\boldsymbol{\nabla} T(\boldsymbol{x}) \cdot \kappa \boldsymbol{x}) {T}(\boldsymbol{x}+\boldsymbol{r})\rangle_{\boldsymbol{x}}  \nonumber \\
&={\xi}_{T}(\boldsymbol{r})+\langle(\boldsymbol{\nabla} T(\boldsymbol{x}+\boldsymbol{r}) \cdot \kappa \boldsymbol{r}) {T}(\boldsymbol{x})\rangle_{\boldsymbol{x}} , 
\label{eq:lensed_corr}
\end{align}
\normalsize
where the last two terms in the penultimate line vary sinusoidally with the polar angle of $\boldsymbol{x}$, and thus average to zero when the expectation value is taken. Simplifying the terms within the expectation value we find that
\begin{equation}
\langle(\boldsymbol{\nabla} T(\boldsymbol{x}+\boldsymbol{r}) \cdot \kappa \boldsymbol{r}) {T}(\boldsymbol{x})\rangle_{\boldsymbol{x}}=r \frac{\partial \xi_{T}}{\partial r} ( \kappa_0 + \gamma_{+,r}) + \frac{\partial \xi_{T}}{\partial \phi} \gamma_{\times,r} \, ,
\end{equation}
where 
\begin{equation}
\gamma_{+,r}+ i \gamma_{\times,r} = (\gamma_{+}+ i \gamma_{\times} ) e^{-2i\phi}
\end{equation}
is the shear in the basis defined by the $\boldsymbol{r}$ direction. The isotropy of the unlensed CMB means that 
$\frac{\partial \xi_{T}}{\partial \phi} =0$, so
\begin{equation}
\langle(\boldsymbol{\nabla} T(\boldsymbol{x}+\boldsymbol{r}) \cdot \kappa \boldsymbol{r}) {T}(\boldsymbol{x})\rangle_{\boldsymbol{x}}=r \frac{\partial \xi_{T}}{\partial r} ( \kappa_0 + \gamma_{+} \cos 2 \phi_r + \gamma_{\times} \sin 2 \phi_r).
\end{equation}
Substituting this into the lensed correlation function expression in eq. (\ref{eq:lensed_corr}), we find that the final expression relating the lensed correlation function to the unlensed correlation function is given by
\begin{equation}
\tilde{\xi}_{T}(\boldsymbol{r})= {\xi}_{T}(\boldsymbol{r})+\frac{\partial \xi_{T}}{\partial \ln r} ( \kappa_0 + \gamma_{+} \cos 2 \phi_r + \gamma_{\times} \sin 2 \phi_r).  
\end{equation}
A similar calculation for $\tilde{\xi}_p(\boldsymbol{r})$, $\tilde{\xi}_m(\boldsymbol{r})$ and $\tilde{\xi}_c(\boldsymbol{r})$ shows that the same relation holds for these lensed correlation functions.

\bibliographystyle{JHEP}	
\bibliography{lensing_refs}{}

\end{document}